%% file: sample-sigconf.tex
\documentclass[sigconf]{acmart}
\usepackage{makecell}
\usepackage{booktabs} 
\usepackage{graphicx}
\usepackage{epsfig}
\usepackage{booktabs} 
\usepackage{footmisc}
\usepackage{threeparttable}
\usepackage{colortbl} 
\usepackage{color}
\usepackage{textcomp}
\usepackage{soul} 
\usepackage{amsmath}
\usepackage{comment}
\usepackage{multirow}
\usepackage{algorithm} 
\usepackage{algpseudocode} 

\usepackage{booktabs}
\usepackage{tablefootnote}
\usepackage{threeparttable}
\usepackage{subfig}
\AtBeginDocument{%
  \providecommand\BibTeX{{%
    \normalfont B\kern-0.5em{\scshape i\kern-0.25em b}\kern-0.8em\TeX}}}

\copyrightyear{2020}
\acmYear{2020}
\setcopyright{acmcopyright}\acmConference[SIGIR '20]{Proceedings of the 43rd International ACM SIGIR Conference on Research and Development in Information Retrieval}{July 25--30, 2020}{Virtual Event, China}
\acmBooktitle{Proceedings of the 43rd International ACM SIGIR Conference on Research and Development in Information Retrieval (SIGIR '20), July 25--30, 2020, Virtual Event, China}
\acmPrice{15.00}
\acmDOI{10.1145/3397271.3401055}
\acmISBN{978-1-4503-8016-4/20/07}

\begin{document}

\title{Detecting User Community in Sparse Domain via Cross-Graph Pairwise Learning}

\author{Zheng Gao}
\affiliation{%
	\institution{Indiana University Bloomington}
}
\email{gao27@indiana.edu}

\author{Hongsong Li}
\affiliation{%
	\institution{Alibaba Group}
}
\email{hongsong.lhs@alibaba-inc.com}

\author{Zhuoren Jiang}
\affiliation{%
	\institution{Zhejiang University}
}
\email{jiangzhuoren@zju.edu.cn}

\author{Xiaozhong Liu}
\affiliation{%
	\institution{Indiana University Bloomington}
}
\email{liu237@indiana.edu}


\begin{abstract}
Cyberspace hosts abundant interactions between users and different kinds of objects, and their relations are often encapsulated as bipartite graphs. Detecting user community in such heterogeneous graphs is an essential task to uncover user information needs and to further enhance recommendation performance. While several main cyber domains carrying high-quality graphs, unfortunately, most others can be quite sparse. However, as users may appear in multiple domains (graphs), their high-quality activities in the main domains can supply community detection in the sparse ones, e.g., user behaviors on Google can help thousands of applications to locate his/her local community when s/he uses Google ID to login those applications.
In this paper, our model, Pairwise Cross-graph Community Detection (PCCD), is proposed to cope with the sparse graph problem by involving external graph knowledge to learn user pairwise community closeness instead of detecting direct communities. Particularly in our model, to avoid taking excessive propagated information, a two-level filtering module is utilized to select the most informative connections through both community and node level filters. Subsequently, a Community Recurrent Unit (CRU) is designed to estimate pairwise user community closeness. Extensive experiments on two real-world graph datasets validate our model against several strong alternatives. Supplementary experiments also validate its robustness on graphs with varied sparsity scales.

\end{abstract}



\maketitle
\input{content/intro.tex}
\input{content/method.tex}

\input{content/experiement.tex}
\input{content/review.tex}
\input{content/conclusion.tex}

\bibliographystyle{ACM-Reference-Format}
\bibliography{acmart} 

\end{document}

%% file: content/intro.tex
\section{Introduction}
\begin{figure}  
 \centering
 \includegraphics[width=1\columnwidth]{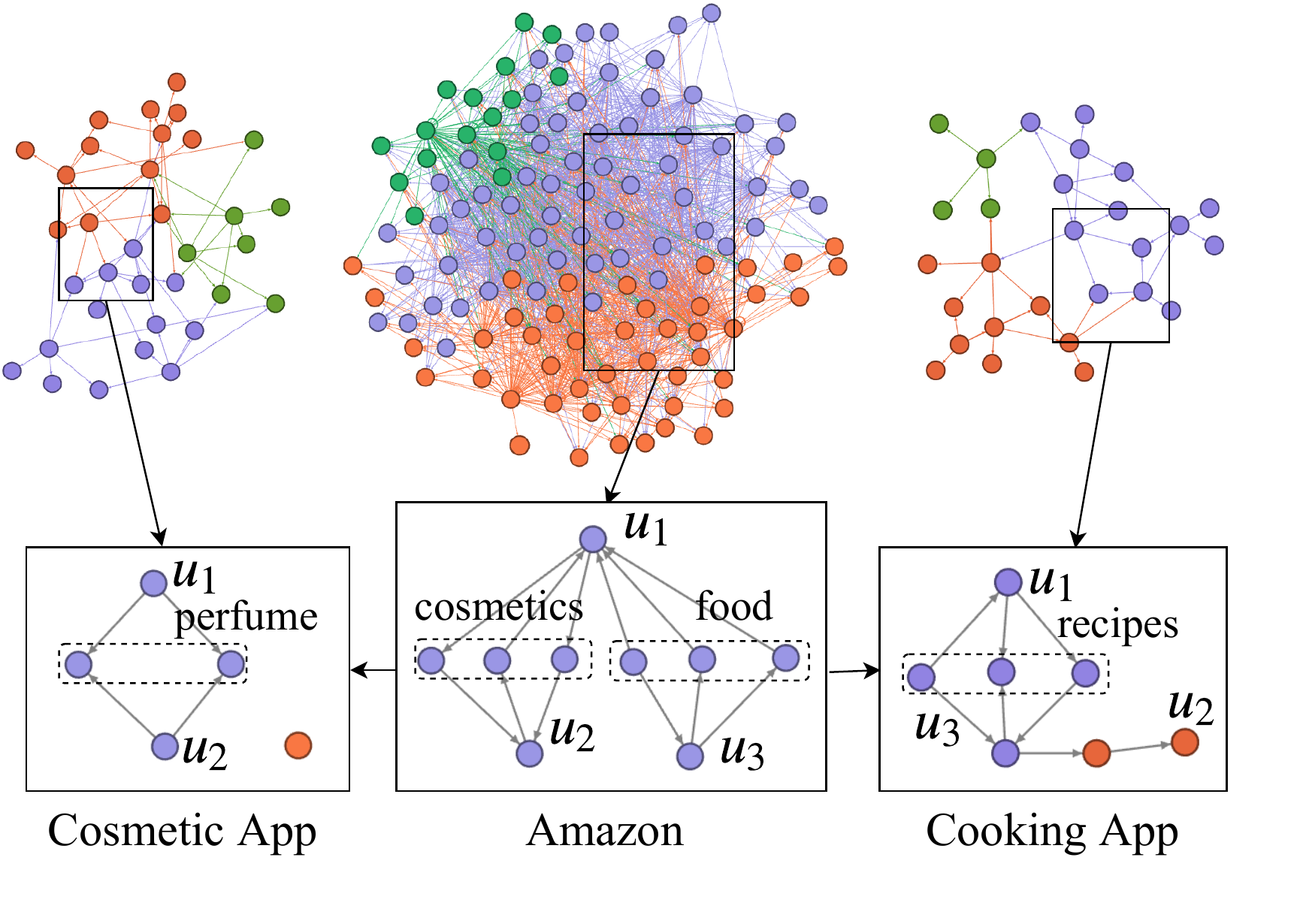}
 \caption{The Amazon graph is a main shopping graph while the Cosmetic App and the Cooking App are two small sparse graphs. Those graphs share mutual users in which three of them are selected to demonstrate their behaviors. Node colors indicate their related communities. }
 \label{fig:example}
  \vspace{-1em}
\end{figure}

Community detection is an essential task for cyberspace mining, which has been successfully employed to explore users’ resemblance for retrieval/recommendation enhancement and user behavior analysis. Taking social media and e-commerce as examples, the complex, and often heterogeneous, relations among users and other objects, e.g., products, reviews, and messages, can be encapsulated as bipartite graphs, and the topology can help to synthesize and represent users with a coarser and broader view.

While a graph is well-connected, conventional methods, e.g., modularity-based approach \cite{haq2019community}, spectral approach \cite{tang2019kernel}, dynamic approach \cite{emmons2019map} and deep learning approach \cite{sun2019vgraph}, are able to estimate the internal/external connectivity and generate high-quality communities directly on nodes \cite{fortunato2016community}.

For a vulnerable graph with sparse connectivity, however, prior community detection algorithms can hardly probe enough information to optimize the community structure. Unfortunately, in the real cyberspace this can be a very common problem, i.e., while a handful of giant players (like Google, Facebook and Amazon) maintaining high-quality graphs, thousands of apps are suffering from graph cold start problem. If many users are isolated because of data sparseness, we can hardly tell any community information. 

In order to cope with this challenge, in this paper, we propose a novel research problem – Cross Graph Community Detection. The idea is based on the fact that an increasing number of small apps are utilizing the user identity information inherited from giant providers, i.e., users can easily login a large number of new apps by using Facebook and Google ID. In such ecosystem, the main large graph can provide critical information to enlighten the community detection on many small sparse graphs. Figure \ref{fig:example} depicts an example, where the Cooking or Cosmetic Apps inherits important topological information from the main Amazon graph for their enhanced community detection. 

Note that, while the small sparse graphs can engage with a local field (like cooking or cosmetic in the example), the main graph can be quite comprehensive and noisy. As Figure \ref{fig:example} shows, not all the connections in Amazon (shopping graph) can be equally important for the two candidate app graphs. Three mutual users are selected where $u_1$ and $u_2$ mainly share similar shopping interests on cosmetics and $u_1$ and $u_3$ mainly share similar shopping interests on food products in Amazon. Then, with deliberate propagation from main graph, in the Cosmetic graph, $u_1$ and $u_2$ have a better chance to be grouped together, while $u_1$ and $u_3$ are more likely to be assigned the same community ID in the Cooking graph. Therefore, the proposed model should be able to differentiate various kinds of information from the main graph for each candidate sparse graph to enhance its local community detection performance.

As another challenge, small sparse graphs often suffer from training data insufficiency, e.g., the limited connections in these graphs can hardly tell the community residency information. In this paper, we employed a novel data augmentation approach - cross graph pairwise learning. Given a candidate user and an associated user triplet, the proposed model can detection the community closeness superiority by leveraging main graph and the sparse graph simultaneously. Moreover, the proposed pairwise learning method can reduce noisy information by taking care of graph local structure. Theoretically, we can offer at most $\mathcal{O}(N^{3})$ user triplets to learn graph community structure while conventional community detection methods by default can only be applied on $\mathcal{O}(N)$ users  ($N$ is the number of users in the sparse graph).


Inspired by aforementioned discussion, we propose an innovative \textit{Pairwise Cross-graph Community Detection} (PCCD) model for enhanced sparse graph user community detection. Specifically, given user $u_i$ and its associated triplet $\langle u_{i},u_{j},u_{k}\rangle$, we aim to predict their pairwise community relationship, e.g., compared with user $u_{k}$, user $u_{j}$ should have closer, similar or farther community closeness to user $u_i$. The contribution of this paper is fourfold: 
\begin{itemize}
    \item We propose a novel problem - Cross Graph Community Detection, which can be critical for thousands of small businesses or services if they enable external user account login.

    \item Unlike conventional community detection methods, we explore community structure from a pairwise viewpoint. In this way, we can efficiently deal with bipartite graph information and solve cold-start problem for users with no behaviors in sparse graphs.
    
    \item The proposed model is trained in an end-to-end manner where a two-level filtering module locates the most relevant information to propagate between graphs. A Community Recurrent Unit (CRU) subsequently learns user community distribution from the filtered information.
    
    \item Extensive experiments on two real-world cross-graph datasets validate our model superiority. We also evaluate our model robustness on graphs with varied sparsity scales and many other supplementary studies. 
\end{itemize} 

%% file: content/method.tex
\section{Method}
\begin{figure*}  
 \centering
 \includegraphics[width=1\textwidth]{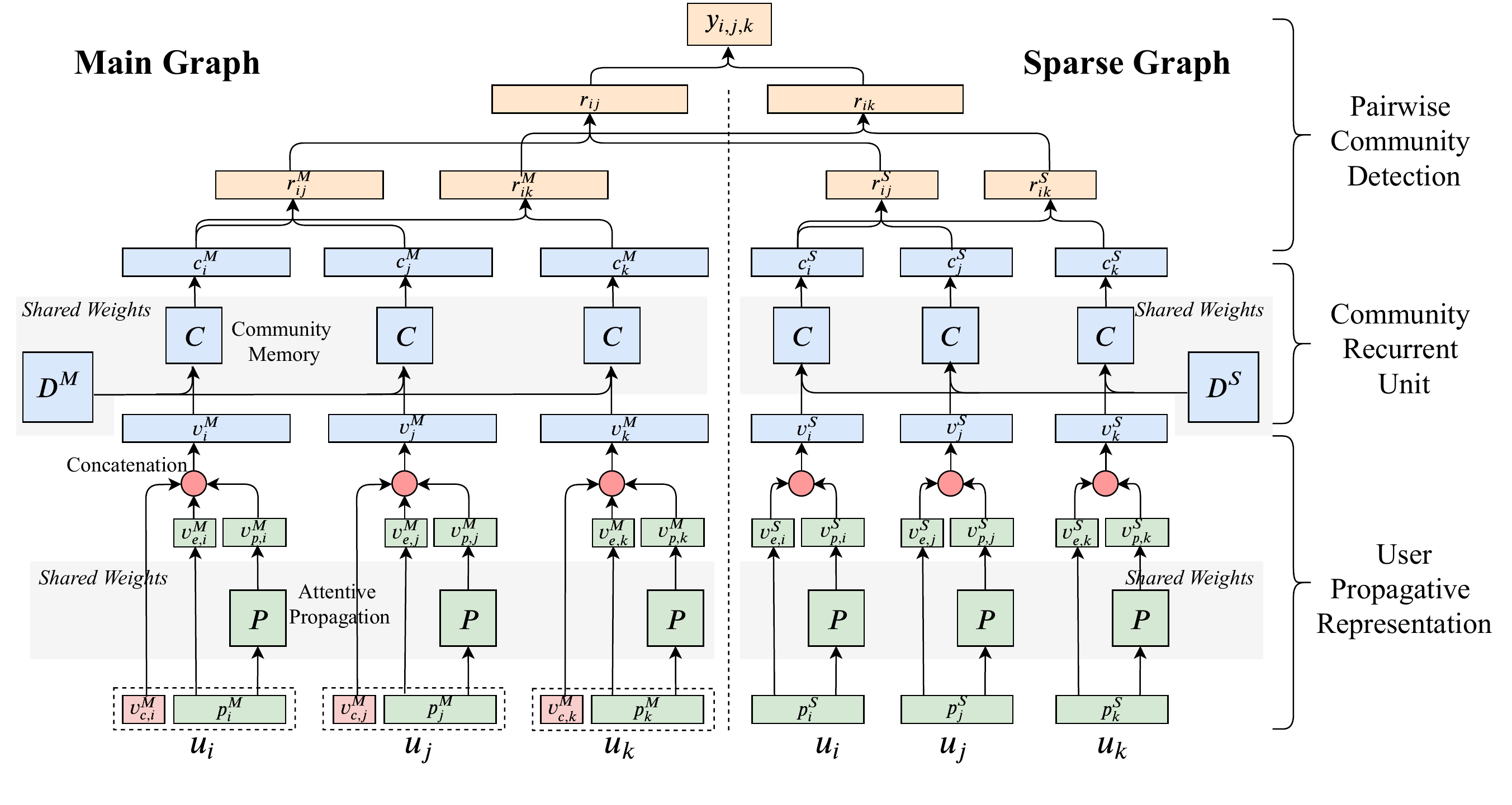}
 \caption{The overall architecture of our proposed PCCD model. It contains three major modules mentioned in the right part of the figure. Each training instance is a mutual user triplet $\langle u_i,u_j,u_k \rangle$. Compared with the sparse graph, the main graph involves raw user community as part of model input.}
 
 \label{fig:pipeline}
 \vspace{-1em} 
\end{figure*}
\subsection{Task Overview} \label{sc:to}

As aforementioned, conventional methods suffer from graph sparseness problem. In this study, 
we propose a Pairwise Cross-graph Community Detection (PCCD) model that particularly aims at detecting user pairwise community closeness in sparse graphs by involving cross-graph techniques. 

Particularly, a well connected graph is called "main graph" in this paper, corresponding to the targeted sparse graph. In general, as users may visit multiple cyber domains within a short time period, these mutual users co-occurred in both the main and sparse graph are taken as the bridge to connect the two graphs. Therefore, the relevant information from the main graph can be propagated to the sparse graph to support its user community detection. 

Specifically, our proposed model (showed in Figure \ref{fig:pipeline}) is trained on mutual user triplets to learn three types of pairwise community relationship including ``\textit{similar}'', ``\textit{closer}'' and ``\textit{farther}''. The goal of this study can be reformulated as the following pairwise learning task: Given a sparse graph $S$ and a main graph $M$ where $N^C$ denotes their mutual user collection, for a mutual user triplet $\langle u_i,u_j,u_k \rangle  \in N^C$ (\textbf{Model Input}),  we aim to predict the relationship of its pairwise community closeness in graph $S$ (\textbf{Model Output}). In this paper, ``\textit{similar}'' relationship means that $u_j$ and $u_k$ are either in the same community or different communities with $u_i$. ``\textit{closer}'' relationship means that $u_j$ and $u_i$ are in the same community, while $u_k$ and $u_i$ are in different communities. ``\textit{farther}'' relationship means that $u_j$ and $u_i$ are in different communities, while $u_k$ and $u_i$ are in the same community.

For a mutual user triplet, our proposed model detects communities firstly on separate graphs to obtain all user raw community representations (Section \ref{sc:rcr}). Their propagative representations are learned through a community-level and a node-level filter (Section \ref{sc:upr}). Both representations are jointly utilized for predicting user community affiliation in each graph via a Community Recurrent Unit (Section \ref{sc:cru}). In the end, we integrate the community affiliation distributions to identify the pairwise community relationship of the triplet (Section \ref{sc:pcd}). Particular training strategies are introduced in the end (Section \ref{sc:mt}).

Note that even though our model is only trained on mutual users, it is still functional to detect pairwise community closeness on those users solely appeared in either the main or sparse graph. Further experiments in Section \ref{sc:ua} will verify its effectiveness on different types of users.

\subsection{Raw Community Representation} \label{sc:rcr}

As the user behavior is enriched in the main graph $M$, detecting user communities in it would offer auxiliary community features to potentially enhance our model performance. Therefore, we decide to involve user communities in the main graph as a part of model input. The same information from the sparse graph is intentionally omitted because its sparse graph structure is not able to offer reliable community partition results.

Considering all possible community detection methods listed in Section \ref{sc:baseline}, Infomap \cite{kheirkhahzadeh2016efficient} is empirically selected to detect user communities in the main graph $M$. Infomap method simulates a random walker wandering on the graph for $m$ steps and indexes his random walk path via a two-level codebook. Its final goal aims to generate a community partition with the minimum random walk description length, which is calculated as follows:

\begin{equation} 
\textit{$L(\pi)$} = \sum_{i}^{m}q_{\curvearrowright}^{i}H(\mathcal{Q})+\sum_{i=1}^{m} \textit{$p_{\circlearrowright}^{i}$}H(\textit{$\mathcal{P}^{i}$}) 
\end{equation}
where \textit{$L(\pi)$} is the description length for a random walker under current community partition $\pi$. $q_{\curvearrowright}^{i}$ and $p_{\circlearrowright}^{i}$ are the jumping rates between and within the $i_{th}$ community in each step. $H(\mathcal{Q})$ is the frequency-weighted average length of codewords in the global index codebook and $H(\mathcal{P}^{i})$ is frequency-weighted average length of codewords in the $i_{th}$ community codebook.

In the end, given a user $u_i$, we obtain its one-hot community representation $v_{c,i}^M$ in the main graph $M$, which is regarded as part of user representations for the model input.

\subsection{User Propagative Representation}\label{sc:upr}

\begin{figure}  
	\centering
	\includegraphics[width=\columnwidth]{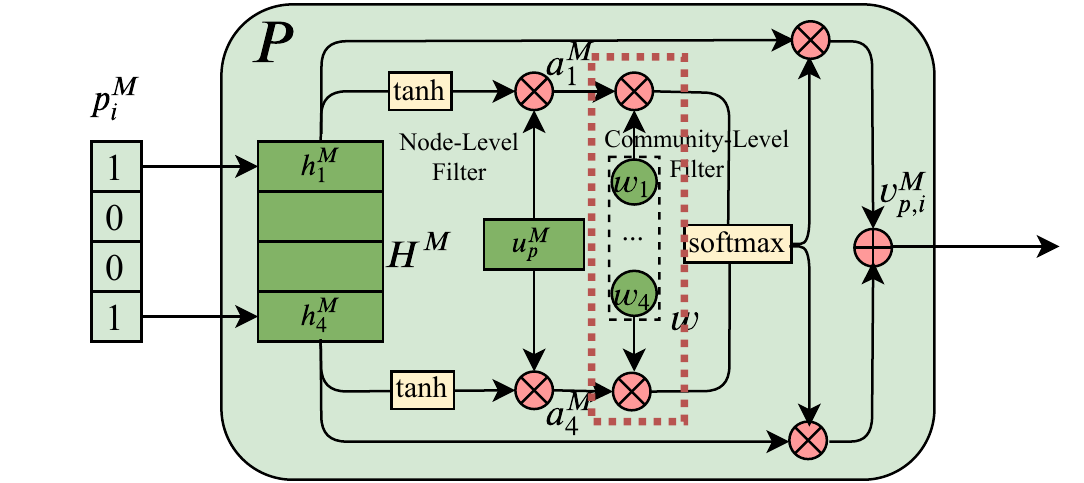}
	\caption{The two-level filter to support user propagative representation estimation in the main graph $M$. Sparse graph $S$ does not contain the community-level filter ( wrapped in the red rectangle).}
	\label{fig:propagation}
	\vspace{-1em} 
\end{figure}

To better convey the mutual user insights in both main and sparse graph, their representations are learned from both direct transformation and weighted information propagation. The optimization processes are similar in the main and sparse graph. In the beginning, a user $u_i$ can be represented as a multi-hot embedding $p_{i}^{M}$ in the main graph $M$ (left part in Figure \ref{fig:propagation}) and $p_{i}^{S}$ in the sparse graph $S$. Each dimension in the embedding refers to a unique object with which the user connects in the related graph. 

The direct transformation from user multi-hot embedding learns a dense vector via one hidden layer so as to retrieve the compressed  information from user connected objects directly, which is calculated as follows:

\begin{equation}
    v_{e,i}^{G} = {\rm tanh}(W_e^Gp_i^G + b_e^G) 
\end{equation}

where $G \in \{M, S\}$ denotes either the main or the sparse graph. $W_e^G$ and $b_e^G$ denote the weight matrix and bias respectively. $v_{e,i}^{G}$ is the learned direct transformation representation for user $u_i$.

On the other hand, from information propagation perspective, each object carries information, which can be propagated to its connected users as a type of representation. However, as such information are noisy, not all of them are equally important to users. Therefore, a filtering module is proposed to automatically select appropriate information to propagate from both community level and node level.  The overall flow of the filtering module is illustrated in Figure \ref{fig:propagation}.

\textbf{Community-Level Filter:} As the raw detected communities in the sparse graph is not reliable because of graph sparseness, we only apply the community-level filter in the main graph $M$. According to the raw community partition $\pi$ calculated in Section \ref{sc:rcr}, each object node $n_i$ is assigned to a community $\pi(n_i)$. The community-level filter propagate object information in a coarser manner by considering its community importance weight $w_{\pi(n_i)}$. Throughout this filter, we aim to learn a weight vector $w$ where each dimension of $w$ denotes the importance score of a related community.

\textbf{Node-Level Filter:} Node-level filter assesses the propagated information with a fine resolution. In the main graph $M$ ( a user-object bipartite graph), we aim to learn object representations $H^M = \{h_1^M,...,h_n^M\}$ where $h_n^M$ denotes the $n_{th}$ object representation. Similarly, $H^S$ denotes the object representations in the sparse graph $S$. The weight of the $j_{th}$ object node decides the amount of its information propagating to its connected user nodes, which is calculated in an attentive means: 

\begin{equation} \label{eq:np}
a_{j}^{G} = (u_p^{G})^{\intercal} {\rm tanh}(W^{G}_p h_{j}^{G} +b_p^{G})
\end{equation}

where $G \in\{M,S\}$ denotes either the main or the sparse graph. $u^{G}_p$ denotes the project vector in the related graph to calculate object attentive weights. $W_p^{G}$ and $b_p^{G}$ denote the corresponding weights and bias, respectively. 

We combine the community-level and node-level weights together to quantify the information propagation in the main graph $M$. While only the node-level weights is considered in the sparse graph $S$. Normalized by the ${\rm softmax}(\cdot)$ function, the related representation of  user $u_i$ in both graphs are calculated as follows:
\begin{equation}\label{eq:pv}
\begin{aligned} 
& v_{p,i}^{M} = \sum_{n_j \in \mathcal{N}^{M}(u_i)} {\rm softmax}(a_{j}^{M} w_{\pi(n_j)})  h_{j}^{M} \\
& v_{p,i}^{S} = \sum_{n_j \in \mathcal{N}^{S}(u_i)}{\rm softmax}(a_{j}^{S}) h_{j}^{S} 
\end{aligned} 
\end{equation}

Both $v_{p,i}^{M}$ and $v_{p,i}^{S}$ are the weighted sum of the neighbour object representations. $\mathcal{N}^{S}(n_i)$ and $\mathcal{N}^{M}(n_i)$ denote user connected objects in both graphs. 

Finally, given a user $u_i$, its raw community representation, direct transformation representation and two-level filtered representation construct its propagative representation in the main graph $M$. While its direct transformation representation and node-level filtered representation construct its propagative representation in the sparse graph $S$. To avoid gradient vanishing, a ${\rm batch\_norm(\cdot)}$ function is applied on top of the concatenated representations in both graph:
\begin{equation} \label{eq:nv}
\begin{aligned} 
& v_{i}^{M} = {\rm batch\_norm}([v_{c,i}^{M},v_{e,i}^{M},v_{p,i}^{M}]) \\
& v_{i}^{S} = {\rm batch\_norm}( [v_{e,i}^{S},v_{p,i}^{S}])
\end{aligned} 
\end{equation} 

\subsection{Community Recurrent Unit}\label{sc:cru}
\begin{figure}  
	\centering
	\includegraphics[width=1.2\columnwidth]{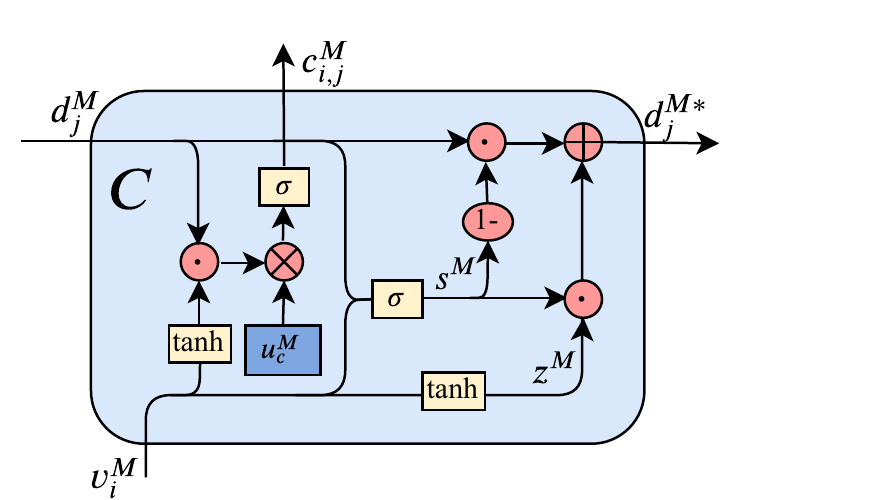}
	\caption{The flow of Community Recurrent Unit in the main graph $M$. The left part refers to the community affiliation gate and the right part is the community update gate.}
	\label{fig:cru}
	\vspace{-1em} 
\end{figure}
After previous steps, each user of the mutual user triplet $\langle u_{i},u_{j},u_{k}\rangle$ is associated with a propagative representation. In this section, its corresponding community affiliation scores $c_{i}^{G}$ are further calculated in related graphs $G \in\{M,S\}$ through a designed Community Recurrent Unit (CRU), which is showed in Figure \ref{fig:cru}. The CRU contains an affiliation gate to calculate the user affiliation score for each community and an update gate to update community self representations. Within this unit, a community memory $D^{G} = \{d_1^G,...,d_K^G\}$ is designated to store $K$ community representations. Particularly, community representation is required to be with the same dimension as user propagative representation in both graphs.

\textbf{Community Affiliation Gate:} The affiliation gate helps to generate the affiliation score of a user $u_i$ towards each community in both graphs, which forms a $K$-dimensional vector $c_{i}^{G} = \{c_{i,1}^{G},...,c_{i,K}^{G}\}$ where $G \in \{M, S\}$ . The user $u_i$'s affiliation score $c_{i,j}^G$ towards the $j_{th}$ community in the graph $G$ is calculated as follows:

\begin{equation}\label{eq:dg}
	c_{i,j}^G = {\rm  \sigma}((u_{c}^{G})^{\intercal}({\rm  tanh}(v_{i}^{G})*d_j^G))
\end{equation}

The dot product between transformed user propagative representation ${\rm  tanh}(v_{i}^{G})$ and the $j_{th}$ community representation $ d_j^G$ indicates their potential correlation, which further  turns to a scalar affiliation score $c_{i,j}^G$ between 0 to 1 with the help of a projection vector $u_{c}^{G}$ and normalization function $\sigma(\cdot)$.
 
\textbf{Community Update Gate: } \label{sc:cug} When calculating $u_i$'s community affiliation, its information can help to update community representations in return. The updated representation is relied on both previous step community representation and current user propagative representation. Therefore, to better embrace the two representations, we use a delicate RNN variant, where the update process is calculated as follows:

\begin{equation}\label{eq:up}
\begin{aligned}
&  s^G = \sigma(W_s^G[v_i^{G},d_j^{G}] + b_s^G) \\
& z^G = {\rm tanh}(W_z^Gv_i^G+b_z^G)\\
& d_j^{G*} = (1-s^G)*d_j^{G} + s^G*z^G
\end{aligned}
\end{equation}

where $G \in \{M,S\}$. $s^G$ denotes the update rate and $z^G$ is transformed user information to be updated in current community.  $d_j^{G*}$ denotes the updated representation of the $j_{th}$ community in graph $G$, which is the weighted sum on previous community and current user representations. $W_s^G$ and $W_z^G,$ denote related weight matrices, while $b_s^G$ and $b_z^G,$ denote related biases. 

\textbf{Community Constraint: }
To obtain the communities with less overlapping information in graph $G \in \{M,S\}$, the community representations ideally should be independent with each other. In our model, cosine similarity ${\rm cos}(\cdot)$ is taken as the criteria to measure the community similarity where higher score indicates stronger correlation. The averaged cosine similarities of all possible community representation pairs is calculated as a community loss to minimize:

\begin{equation}\label{eq:cc}
	\mathcal{L}_{c}^{G} = \frac{1}{2K^2}\sum_{i,j}{\rm cos}(d_i^{G},d_j^{G})
\end{equation}
where $K$ is the number of communities in each graph.

\subsection{Pairwise Community Detection}\label{sc:pcd}

Similar to RankNet \cite{burges2010ranknet}, given a mutual user triplet  $\langle u_i,u_j,u_k \rangle$, as three types of pairwise label are considered including ``\textit{closer}'', ``\textit{similar}'' and ``\textit{farther}'', label $y_{i,j,k}$  is calculated as follows:
\begin{equation}\label{eq:y}
	y_{i,j,k} = \frac{1}{2}(1+S_{jk})
\end{equation}
In terms of the community closeness for $u_i$, if $u_j$ is closer than $u_k$, $S_{jk} = 1$; if $u_j$ is farther than $u_k$, $S_{jk} = -1$; if $u_j$ and $u_k$ are similar, $S_{jk} = 0$. In this way, we convert the pairwise ranking task to a regression task. 

To optimize this task, first of all, we calculate the correlation representation $r^{G}_{ij}$ between $u_j$ and $u_i$ in single graph $G \in \{M,S\}$:

\begin{equation}\label{eq:gr}
	r^{G}_{ij} = W^{G}_{r}(c_{i}^{G} - c_{j}^{G}) +  b^{G}_{r} 
\end{equation}

where $ W^{G}_{r}$ and $ b^{G}_{r}$ are related weight and bias, respectively. 

To construct the information from both graphs, we concatenate the correlation representations from both graphs:
\begin{equation} \label{eq:cr}
	r_{ij} = {\rm tanh}([r^{S}_{ij},r^{M}_{ij}])
\end{equation}
Similarly, the cross-graph correlation representation  between $u_i$ and $u_k$ is calculated as $r_{ik}$. 

After that, we predict the community relationship between $u_j$ and $u_k$ towards $u_i$ as follows:
\begin{equation} \label{eq:yp}
\hat{y}_{i,j,k} = \sigma(W_{o}(r_{ij} - r_{ik}) +  b_{o})
\end{equation}

where $W_{o}$ and $b_{o}$ are the related weight and bias. $\sigma(\cdot)$ denotes the sigmoid activation function. In the end, the final loss $\mathcal{L}_{total}$ is the weighted sum of the optimization loss calculated via cross entropy and the community constraint losses where  $\mathcal{L}_{c}^S$ is for the sparse graph and $\mathcal{L}_{c}^M$ is for the main graph:
\begin{equation}\label{eq:loss}
\mathcal{L}_{total} = \underbrace{-y_{i,j,k}log\hat{y}_{i,j,k}-(1-y_{i,j,k})log(1-\hat{y}_{i,j,k})}_{optimization} + \underbrace{\alpha(\mathcal{L}_{c}^S + \mathcal{L}_{c}^M)}_{constraint}
\end{equation}

\subsection{Training Strategy}\label{sc:mt}

Although our model is trained solely on mutual user triplets, it is also able to detect pairwise community closeness among users only appeared in either sparse graph or main graph. Two training tricks are particularly employed in our model to improve model robustness, including shared weights and masked training. 

First, as our model should be insensitive to the sequence of input user triplet, the weight matrices, biases and project vectors in both Section \ref{sc:upr} and Section \ref{sc:cru} should be shared among the three users. Moreover, in  Section \ref{sc:cru}, the community memory $D^G$ where $G \in \{M,S\}$ is updated by taking the average of the three updated community representations (calculated in Eq. \ref{eq:up}) from input triplets.

Second, to alleviate the negative effects of excessive connections in main graph $M$, we employ a masked training strategy by randomly remove a small ratio $\rho$ of connected objects from users in the main graph during each training batch. $\rho = 0$ means users remain their original multi-hot embeddings while $\rho = 1$ means users lose all their connections on objects. 


%% file: content/experiement.tex
\section{Experiment}  \label{sc:exp}
\subsection{Dataset}  \label{sc:dataset}
As one of the largest e-commerce ecosystems, Alibaba owns multiple online businesses, where users could try different services via the same login ID. For this experiment, we employ three services from Alibaba including an online shopping website, a digital news portal and a cooking recipe website. The online shopping website is regarded as the main graph because it is Alibaba core service having the largest number of active users. By contrast, the other two services are regarded as sparse graphs. In the online shopping website, each object refers to a product for sale, and links are user purchase behaviors on products. In the digital news portal, each object refers to a news article, and links are user reads on news. In the cooking recipe website, each object refers to a cooking recipe, and links mean that users follow recipe instructions. 

To prove our model robustness with respect to the graph size, two cross-graph datasets are constructed in different scales, which are showed in Table \ref{tab:dataset}. The SH-NE dataset is the cross-graph dataset constructed by the shopping graph and news graph, which is ten times larger than the SH-CO dataset constructed by the shopping graph and recipe graph. Users are categorized into three different types including \textit{MU} (mutual users appeared in both graphs),  \textit{MO} (users only appeared in the main graph), and  \textit{SO} (users only appeared in the sparse graph). Note that, only mutual user triplets are used for training.
 
\begin{table}[h]
	
	\small
	\centering
	\renewcommand{\tabcolsep}{1.5pt}
	\begin{tabular}{ccccccccc} 
		\toprule
		\multirow{2}{*}{\textbf{Dataset}}& \multirow{2}{*}{\textbf{\#MU}}&\multicolumn{3}{c}{\textbf{Main Graph}}& \multicolumn{3}{c}{\textbf{Sparse Graph}}\\ \cmidrule(lr){3-8}
		&  &\textbf{\#MO}&\textbf{\#Object}& \textbf{\#Link}&\textbf{\#SO}&\textbf{\#Object}& \textbf{\#Link}\\ \midrule
		\textbf{SH-NE}& 105,702 & 392,549 & 135,320 & 17,352,482  & 26,133 & 9,377 & 881,721\\ 
		\textbf{SH-CO}& 1,029  & 1,543  &23,881 & 1,161,293& 719  & 3,739 &147,266\\\bottomrule
	\end{tabular}
	\caption{Statistics of Two cross-graph datasets.}
	\label{tab:dataset}
	\vspace{-1em} 
\end{table}  
 
Both datasets are built by taking a week of user behaviors from 09/20/2018 to 09/26/2018. In this paper, for two sparse graphs, we aim to predict user pairwise community labels (calculated from enriched seven-day user behaviors) using sparse user behaviors (the first four-day user behaviors, arbitrarily defined in this paper).

To construct our ground truth data, we run all baseline models (will be introduced in Section \ref{sc:baseline}) on both seven-day sparse graphs which are assumed to contain sufficient user-product connections during this relatively long time period. Only mutual user communities agreed by all six baselines (except random model) are kept, from which mutual user triplets are subsequently selected as we assume the commonly-agreed communities are reliable. In this paper, we define three types to label pairwise community closeness for a mutual user triplet, including ``\textit{closer}'', ``\textit{similar}'', and ``\textit{farther}'' (detailed definition is in Section \ref{sc:to}). Equal number of triplets are randomly selected for each label type. In total, there are 207,291 triplets in SH-NE dataset and 20,313 triplets in SH-CO dataset for testing propose. 

 Training data generation is the same as ground truth data construction except using first four-day user behaviors. In total, there are 2,072,905 triplets in SH-NE dataset and  203,123 triplets in SH-CO dataset, from which 90\% of the triplets are used for training and 10\% for validation.

In both training and testing process, our model input is user triplets with first four-day behavior information. Their difference is that the pseudo labels for training are generated from four-day user behaviors, while the ground truth labels for testing are generated from whole seven-day user behaviors in sparse graphs.

\subsection{Baselines and Settings} \label{sc:baseline}
As there is neither reliable nor open-sourced studies on pairwise cross-graph community detection, we have no direct comparative models. In this paper, we select one random model and six compromised but state-of-art baselines which are originally designed for whole graph community detection. In all seven baselines, after user communities are detected, we calculate user triplet labels via Eq. \ref{eq:y}. As all baselines are trained on the sparse graph with first four-day user behavior and the main graph, they bring no information loss compared with our model. Here is the details of how these baselines calculate user communities:

\begin{itemize}
	\item \textbf{random}: users are randomly assigned to a given number of communities.
	\item \textbf{Infomap} \cite{kheirkhahzadeh2016efficient}: A random walk based method to detect user communities by minimizing description length of walk paths\footnote{https://github.com/mapequation/infomap}.
	\item \textbf{DNR} \cite{yang2016modularity}: A novel nonlinear reconstruction method to detect communities by adopting deep neural networks on pairwise constraints among graph nodes\footnote{http://yangliang.github.io/code/}.
	\item \textbf{LSH} \cite{chamberlain2018real}: A random walk based method for community detection in large-scale graphs\footnote{https://github.com/melifluos/LSH-community-detection}.
	\item \textbf{node2vec} \cite{grover2016node2vec}: A skip-gram model to generate node emebeddings based on guided random walks on graphs. And K-means method subsequently calculates communities from the generated embeddings\footnote{https://github.com/snap-stanford/snap/\label{fn:snap}}.
	\item \textbf{BigClam} \cite{yang2013overlapping}: A non-negative matrix factorization method to detect overlapping communities in large scale graphs\footref{fn:snap}.
	\item \textbf{WMW} \cite{castrillo2017fast}: A fast heuristic method for hierarchical community detection inspired by agglomerative hierarchical clustering\footnote{https://github.com/eduarc/WMW}.
\end{itemize}

 The result is reported with best hyper-parameters from grid search. Empirically, a mini-batch (size = 200) Adam optimizer is chosen to train our model with learning rate as 0.01. Community number on both graphs is $K = 50$ (Eq. \ref{eq:cc}). Community constraint weight is $\alpha = 0.1$ (Eq. \ref{eq:loss}). Masked training rate in the main graph is $\rho = 0.05$. The model is trained for 1000 epochs in order to get a converged result. 
 
 Model performance is evaluated from both classification and retrieval viewpoints. From classification aspect, Accuracy (ACC), F1-macro (F1) and Matthews Correlation Coefficient (MCC) are reported metrics.
 From retrieval aspect, Mean Reciprocal Rank (MRR), Normalized Discounted Cumulative Gain (NDCG) and Mean Average Precision (MAP) are reported metrics. All of them are fundamental metrics widely used for model evaluation. 
\subsection{Performance Comparison}
\begin{table*}[h]
	
	\centering
	\renewcommand{\tabcolsep}{2.5pt}
	\begin{tabular}{cccccccccccccc} 
		\toprule
		\multirow{2}{*}{ \textbf{Graph}} & 	\multirow{2}{*}{\textbf{Model}}	&\multicolumn{6}{c}{\textbf{SH-NE Dataset}} & \multicolumn{6}{c}{\textbf{SH-CO Dataset}}\\
		\cmidrule(lr){3-14}
		& & \textbf{ACC}& \textbf{F1}& \textbf{MCC}& \textbf{MRR}& \textbf{NDCG}& \textbf{MAP} & \textbf{ACC}& \textbf{F1}& \textbf{MCC} & \textbf{MRR}& \textbf{NDCG}& \textbf{MAP} \\ \midrule 
		\multirow{6}{*}{\textbf{Main }} &random  & 0.3324& 0.1677&-0.0001&0.7719&0.8373&0.7299&0.3302&0.1700&0.0027&0.7842&0.8444&0.7401\\ 
		& Infomap &0.3700& 0.2731&0.1201&0.8334&0.8421&0.7299 &0.3811&0.3723&0.1000&0.7888&0.8419&0.7430\\  
		& DNR &0.4210&0.4022&0.1789&0.8000&0.8366&0.7473&0.5194&0.5122&0.2712&0.8099&0.8773&0.7592\\   
		& LSH &0.3798&0.3813&0.0888&0.8002&0.8433&0.7328&0.4000&0.4123&0.1277&0.7812&0.8329&0.7570\\ 
		& node2vec& 0.4888 &0.5014&0.2521&0.8229&0.8744&0.7832&0.5959&0.5832&0.3895&0.8422&0.8936&0.8222\\   
		& BigClam &0.5555&0.5399&0.3349&0.8421&0.8890&0.8024&0.5301&0.5334&0.3290&0.8335&0.8573&0.8005\\ 
	    & WMW &0.6032&0.6158&0.4111&0.8632&0.8988&0.8256&0.6489&0.6661&0.5001&0.8789&0.9032&0.8499\\ \midrule 
		\multirow{6}{*}{\textbf{Sparse}} &random  & 0.3333& 0.1785&0.0003&0.7815&0.8438&0.7419&0.3301&0.1720&0.0030&0.7801&0.8400&0.7338\\ 
		& Infomap &0.3455& 0.1929&0.0031&0.7843&0.8399&0.7437 &0.3676&0.3211&0.0422&0.7905&0.8446&0.7401\\  
		& DNR &0.4433&0.4407&0.2020&0.8313&0.8678&0.7742&0.5273&0.5289&0.2787&0.8293&0.8765&0.7980\\   
		& LSH &0.3889&0.3917&0.0942&0.8008&0.8528&0.7343&0.4093&0.4177&0.1138&0.7833&0.8438&0.7404\\ 
		& node2vec& 0.4849 &0.4958&0.2467&0.8219&0.8664&0.7774&0.5732&0.5746&0.3898&0.8412&0.8823&0.8146\\   
		& BigClam &0.5277&0.5355&0.3332&0.8441&0.8750&0.8005&0.5282&0.5280&0.3339&0.8298&0.8399&0.8033\\ 
	    & WMW &0.5814&0.5900&0.3988&0.8666&0.9001&0.8210&0.5911&0.6212&0.4347&0.8599&0.8890&0.8133
		\\ \midrule 
		\multirow{7}{*}{\textbf{\makecell{Main \\ + \\ Sparse}}} &random  & 0.3337& 0.1788&-0.0008&0.7815&0.8387&0.7387&0.3291&0.1699&0.0027&0.7739&0.8331&0.7310\\ 
		& Infomap &0.3625& 0.2567&0.0908&0.7864&0.8423&0.7437 &0.3930&0.3988&0.0916&0.7951&0.8488&0.7528\\  
		& DNR &0.4383&0.4204&0.1817&0.8101&0.8599&0.7686&0.5273&0.5322&0.2957&0.8391&0.8812&0.8000\\   
		& LSH &0.3992&0.4052&0.0993&0.8020&0.8538&0.7599&0.4180&0.4237&0.1309&0.8021&0.8539&0.7600\\ 
		& node2vec& 0.5072  &0.5124&0.2611&0.8339&0.8774&0.7943&0.6015&0.6055&0.4045&0.8677&0.9023&0.8324\\   
		& BigClam &0.5580&0.5626&0.3412&0.8511&0.8901&0.8135&0.5398&0.5383&0.3389&0.8475&0.8874&0.8095\\ 
	    & WMW &0.6168&0.6205&0.4281&0.8702&0.9042&0.8353&0.6682&0.6711&0.5041&0.8879&0.9173&0.8561\\ 
		& \textbf{PCCD}& \textbf{ 0.6524*}& \textbf{0.6551*} &  \textbf{0.4808*}& \textbf{0.8802*} & \textbf{0.9116*} & \textbf{0.8470*}& \textbf{0.7740*}& \textbf{0.7758*} &  \textbf{0.6627*}& \textbf{0.9244*} & \textbf{0.9442*} &  \textbf{0.9005*}
		 \\ \bottomrule
	\end{tabular}
	\caption{All model performances for three different graph combinations in two datasets. Symbol `*' highlights the cases where our model significantly beats the best baseline with $p<0.01$.}
	\label{tab:evaluation}
	\vspace{-1.5em} 
\end{table*}

All model performance are evaluated from both classification and retrieval viewpoints. Each baseline method is run on three graphs including the main graph, the sparse graph, and their combined graph. While our model can only be tested by utilizing both graphs given the proposed model structure. Table \ref{tab:evaluation} shows the overall model performance under three graphs. Particularly, we run our model ten times and report the average performance result. To verify our model's superiority, we calculate the performance differences between our model and the best baseline on each metric for all the runs, and apply a pairwise t-test to check whether the performance difference is significant.

Our model output is a predicted score in the range from 0 to 1. Its associated label is assigned to be one of ``\textit{closer}'', ``\textit{similar}'' or ``\textit{farther}'' based on which third the predicted score lies in (associated with label definition in Section \ref{sc:pcd}). For example, a user triplet $\langle u_i,u_j,u_k \rangle$ with score 0.3 will be labelled as ``\textit{farther}'', which means compared with $u_k$, $u_j$ is has a farther community closeness to $u_i$. After that, from classification perspective, ACC, F1 and MCC scores are calculated by comparing predicted labels with the ground truth labels. While from retrieval perspective, MRR, NDCG and MAP are calculated for the pairwise ranking within user triplets.

Table \ref{tab:evaluation} shows that almost all baseline performances in the combined graph achieve the best among three types of graphs. And performances in the main graph are always better than in the sparse graph. It demonstrates that main graph holds beneficial information to reveal user potential communities in sparse graphs. The random baseline is with similar performance among all three graphs, which makes sense because the random community partition is regardless of graph structures. It can be used as the default model to compare with in each graph. For the rest six baselines, random walk based baselines (Infomap and LSH) do not perform well in all three graphs. Their evaluation results are just a bit better than random model performance. Even though solely running Infomap does not perform well, it is the most supportive model for Section \ref{sc:rcr} Raw Community Detection (related empirical experiment comparisons are omitted as it is not the major model contribution). WMW model performs the best among all baselines. Actually, its results are always the best in three graphs of both datasets. However, by taking the pairwise t-test, it statistically proves that our model still significantly outperforms the WMW model in terms of all metrics.

It is also interesting to see that the model evaluation results are consistent in all datasets. For example, DNR model always performs better than LSH model. The only exception is that BigClam performs better than node2vec in SH-NE dataset, but worse in SH-CO dataset. 
 
\subsection{Ablation Study}

In this section, we aim to explore whether all components of our proposed model have positive effect on final model performance and which component has the largest impact. There are six components that can be disassembled from the main model, including Raw Community Representation in the main graph (RCR), Direct Transformation Representation (DTR), Node-level filter (NF), Community-level filter (CF), Community Constraint (CC) and Masked Training (MT). We iteratively remove each component while keep the rest firmed to demonstrate their performance difference compared to the original model. All comparison results are showed in Table \ref{tab:ablation}.
\begin{table}[h]
	
	\centering
	\renewcommand{\tabcolsep}{3pt}
	\begin{tabular}{cccccccc} 
		\toprule
		\textbf{Dataset}& \textbf{Model}	
		& \textbf{ACC}& \textbf{F1}& \textbf{MCC}& \textbf{MRR} & \textbf{NDCG}& \textbf{MAP}\\ \midrule 
		\multirow{5}{*}{ \textbf{SH-NE}} & -- RCR  &-2.61&-2.51&-3.76&-0.73&-0.54&-0.96\\
		& -- DTR &-1.32&-1.49&-2.96&-1.01&-0.14&-0.57\\ 
		& -- NF &-17.31&-17.01	&-25.92&-5.35&-3.88&-6.05\\
		& -- CF &-7.94&-8.48&-11.71&-2.61&-1.92&-3.01\\
		& -- CC &-2.15&-2.05&-3.08&-0.49&-0.35&-0.65\\ 
		& -- MT &-2.43&-2.55&-4.82&-0.78&-0.82&-1.24\\ 
		\midrule 
		\multirow{5}{*}{ \textbf{SH-CO}} & -- RCR &-2.13&-2.45&-3.07&-1.65&-0.58&-0.79  \\  
		& -- DTR &-3.45&-2.89&-1.95&-2.22&-0.43&-0.88\\ 
		& -- NF &-26.13&-25.83&-39.97&-9.81&-7.50&-19.35\\ 
		& -- CF &-21.12&-5.53&-8.45&-2.03&-1.50&-2.50\\
		& -- CC &-4.32&-4.34&-6.57&-1.56&-1.24&-1.91\\
		& -- MT &-1.32&-1.24&-2.66&-0.33&-0.21&-0.45\\ \bottomrule
	\end{tabular}
	\caption{Performance differences on all evaluation metrics between each ablated model and the original model. ``-'' refers to remove related component from our model. Results are scaled in percentage (\%).}
	\label{tab:ablation}
	\vspace{-1.5em} 
\end{table} 

Among the six components, if we remove node-level filter, the performance drop dramatically, even worse than some of baseline models. It indicates that node-level information are excessive and noisy. Without appropriate filtering on the propagated information, it would vitally pollute the quality of user representations. Similarly,  community-level filter also has a huge impact on model performance, meaning filtering propagated information from a coarser view also has a positive effect on learning user representations. By contrast, the two extra representations (RCR and DTR) do not have strong influence in our model. Having these information can slightly improve model performance as they still involve extra information. However, as one-hot embeddings, the amount of new information that RCR can offer is very limited. Similarly, DTR is only one-layer transformation on multi-hot embeddings. The information it can offer is also limited. By adding extra constraint, CC also has a little positive effect to enhance model performance. But as we always set its weight with a small value in training process, its influence is also indifferent. MT has the least impact on model performance as randomly masking user behaviors in the main graph has an overlapped effect with attentive weights calculated from the node-level filter in Section \ref{sc:upr}.

\subsection{User-Type Analysis}
\label{sc:ua}
\begin{table}[h]
	
	\centering
	\renewcommand{\tabcolsep}{3.5pt}
	\begin{tabular}{cccccccc} 
		\toprule
		\textbf{Dataset}& \textbf{User}	
		& \textbf{ACC}& \textbf{F1}& \textbf{MCC}& \textbf{MRR} & \textbf{NDCG}& \textbf{MAP}\\ \midrule 
		\multirow{3}{*}{ \textbf{SH-NE}} & MU & 0.7132 & 0.7133 & 0.6012&  0.9112 & 0.9324 & 0.8704  \\   
	    & MO & 0.6356 & 0.6477 & 0.4780&  0.8739 & 0.9002 & 0.8418 \\
	    & SO & 0.6009 & 0.6117 & 0.4324&  0.8600 & 0.8988 & 0.8393 \\
		\midrule 
		\multirow{3}{*}{ \textbf{SH-CO}} & MU & 0.7852& 0.7780& 0.6724& 0.9300& 0.9474& 0.9015 \\   
	    & MO & 0.7334& 0.7565& 0.6601& 0.9012& 0.9400& 0.8874\\
	    & SO & 0.6823& 0.6915& 0.6844& 0.8990& 0.9222& 0.8789 \\ \bottomrule
	\end{tabular}
	\caption{Three types of user performance in two datasets.}
	\label{tab:ua}
	\vspace{-1.5em} 
\end{table} 

As aforementioned in Section \ref{sc:dataset}, there are three types of users in our cross-graph datasets including \textit{MU} (mutual users appeared in both graphs),  \textit{MO} (users only appeared in the main graph), and  \textit{SO} (users only appeared in the sparse graph). Unlike our training data which is constructed only with mutual users, our testing data can be with all types of users. To examine our model performance on each type of users, each single-type user user triplets are screened with five thousand instances, i.e., an eligible user triplet only contains three \textit{MO} type users. In fact, the performance on \textit{MO} user triplets reflects the capability of our model to solve cold-start problem as these users have no behaviors in the sparse graphs. The result showed in Table \ref{tab:ua} demonstrates that the label of \textit{MU} triplets can be best identified, which makes sense as the information of mutual users come from both graphs. While performance results on \textit{MO} are better than \textit{SO} user triplets, which might because that users are likely to have more enriched behaviors in the main graph compared with the sparse graph. Even that, the performance of our model on \textit{SO} user triplets is still better than most of baselines running on the combined graph.

\subsection{Graph Sparsity Influence}
\begin{figure}
	\centering
	\subfloat[ACC]{\includegraphics[width = 1.6in]{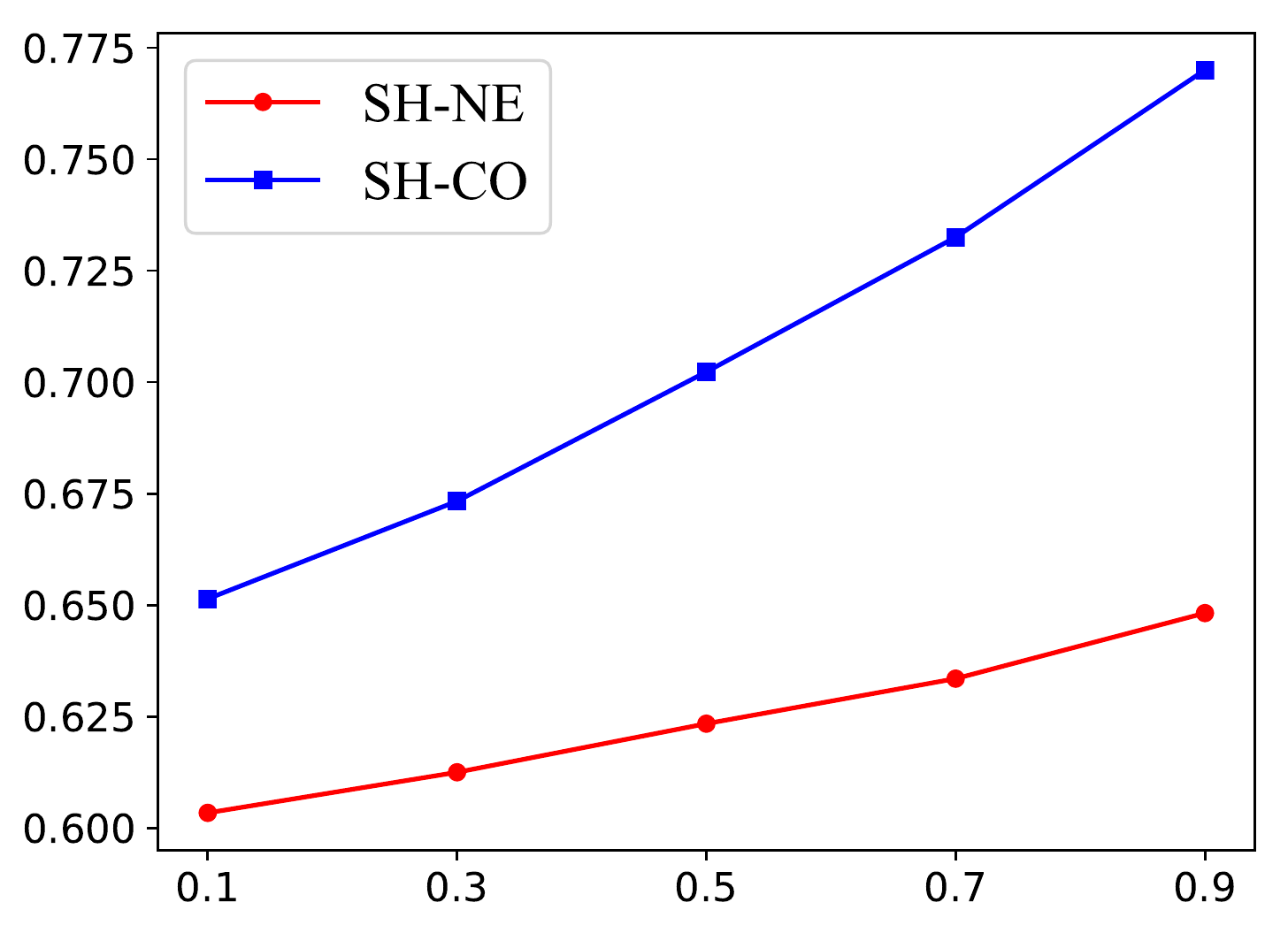}} \hspace{0.1in}
	\subfloat[F1]{\includegraphics[width = 1.6in]{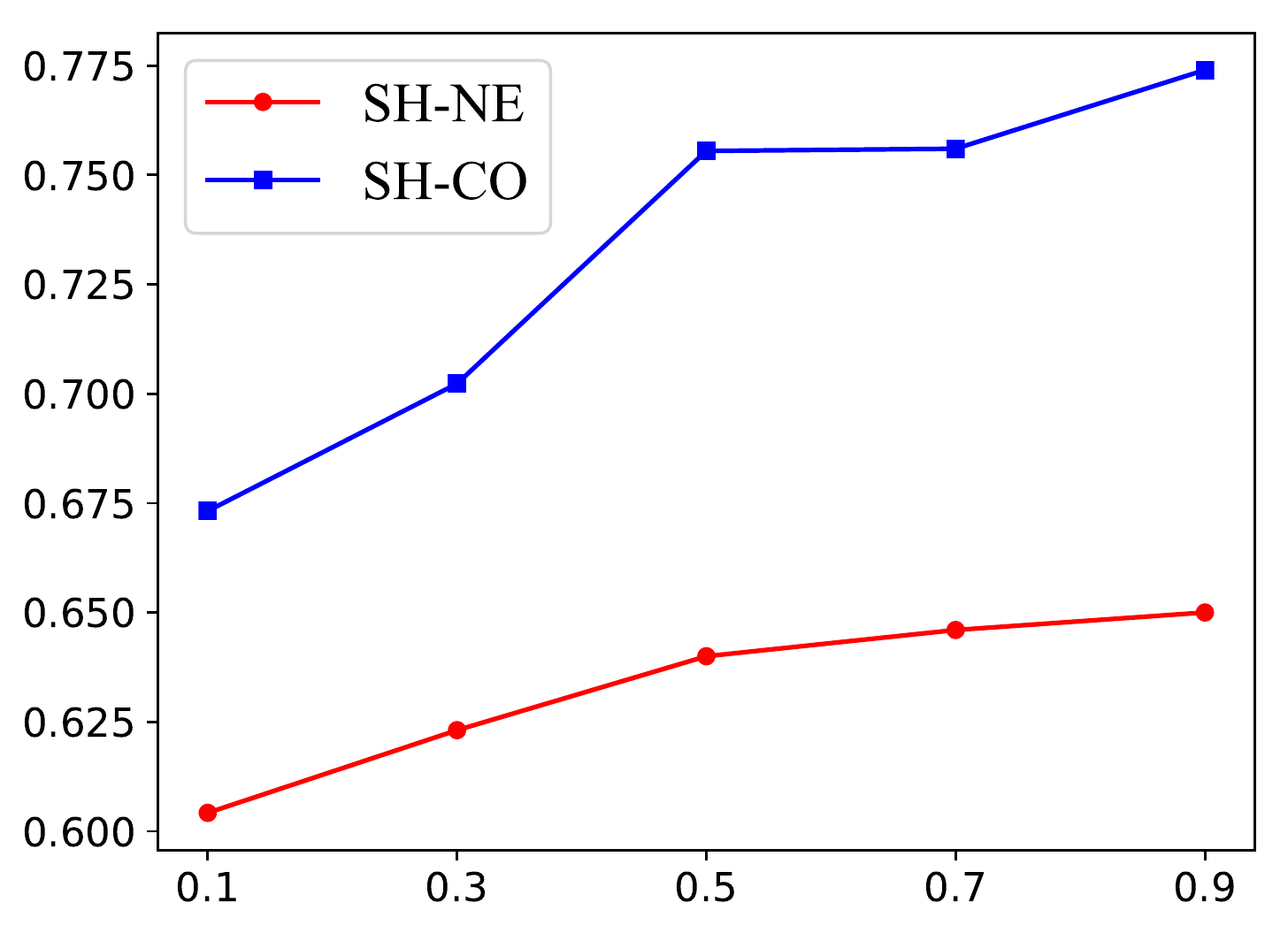}} \vspace{-0.5em} \\
	\subfloat[MCC]{\includegraphics[width = 1.6in]{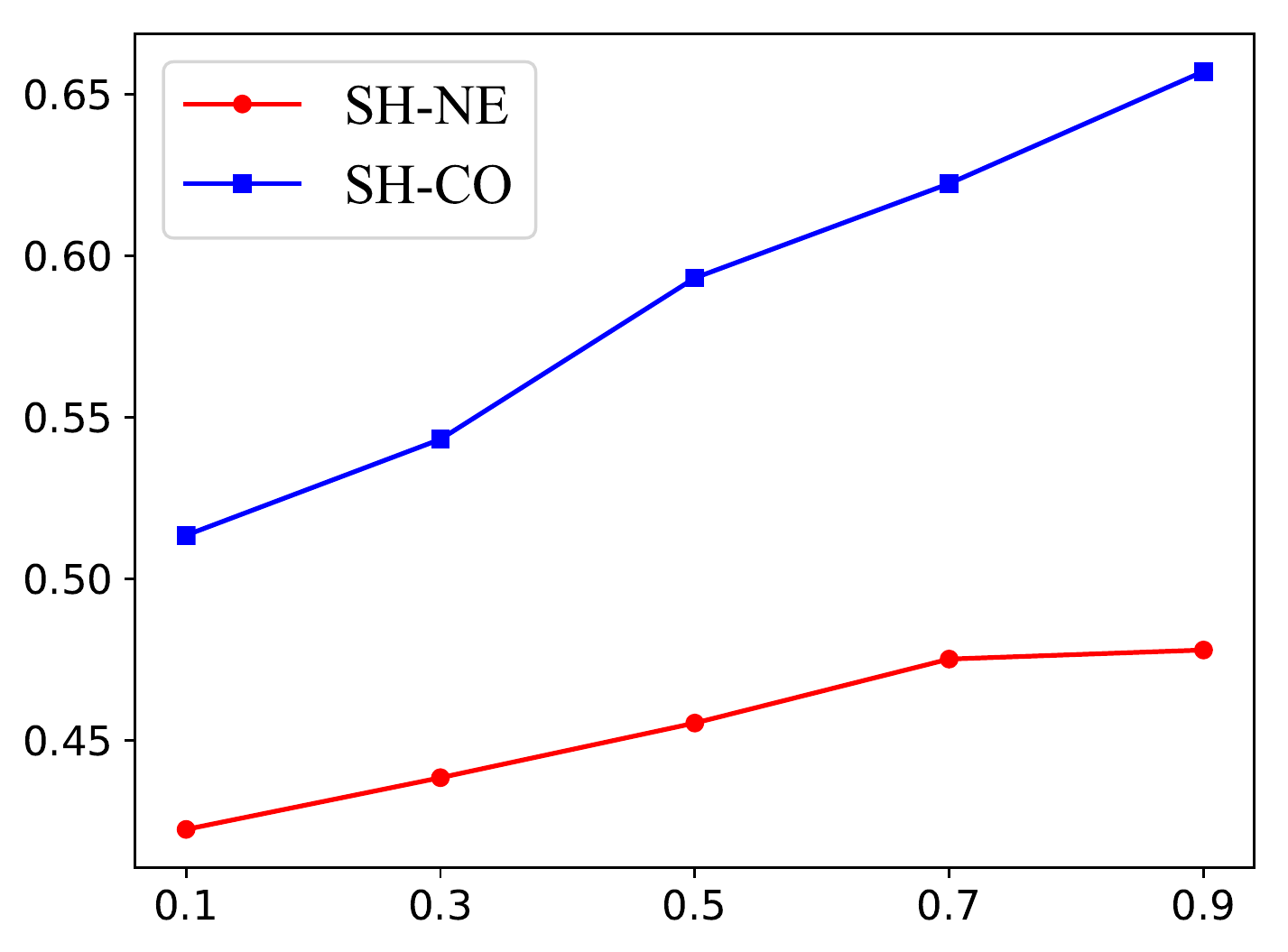}}
	\hspace{0.1in}
	\subfloat[MRR]{\includegraphics[width = 1.6in]{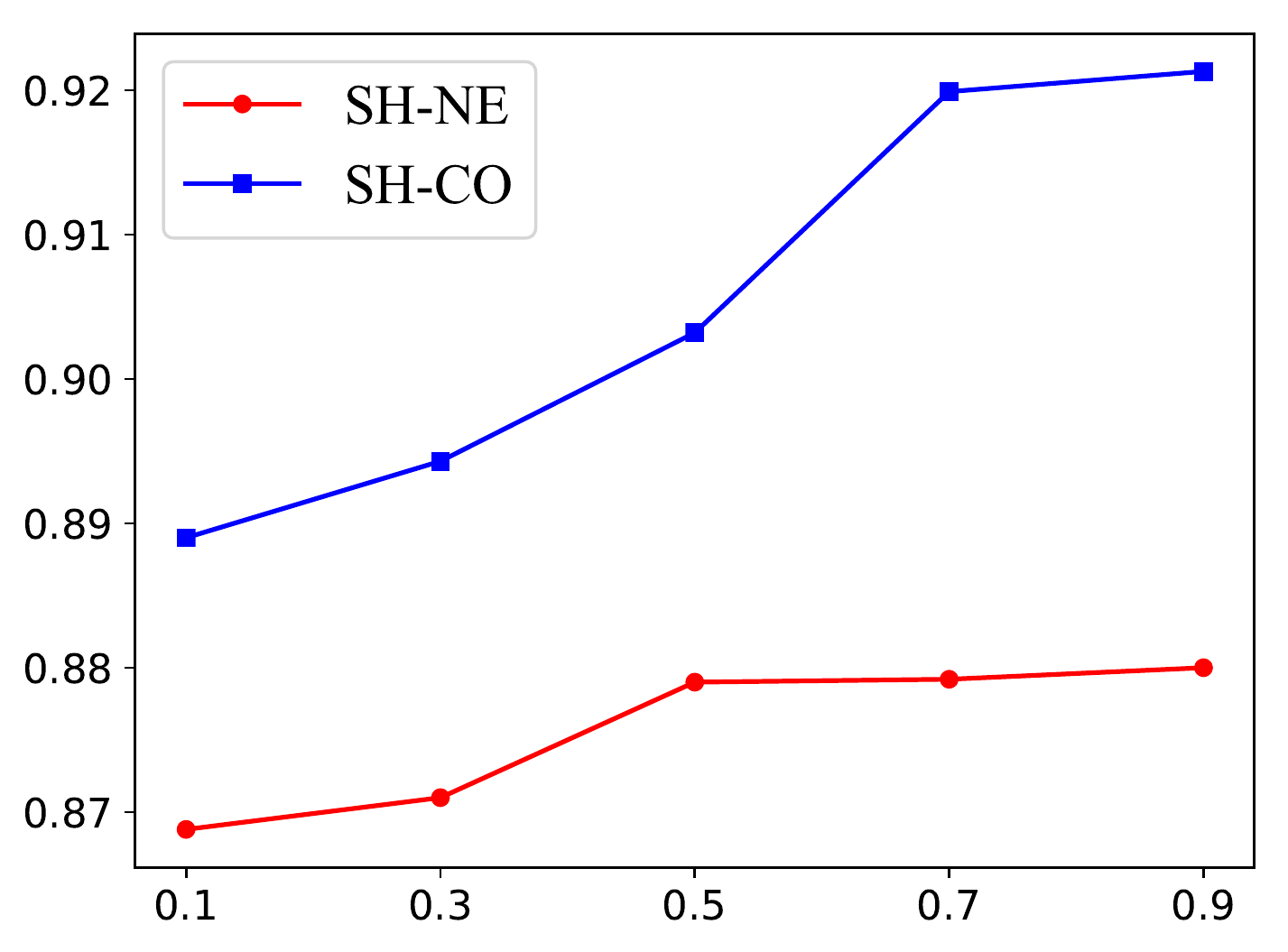}} \vspace{-0.5em}\\ 
	\subfloat[NDCG]{\includegraphics[width = 1.6in]{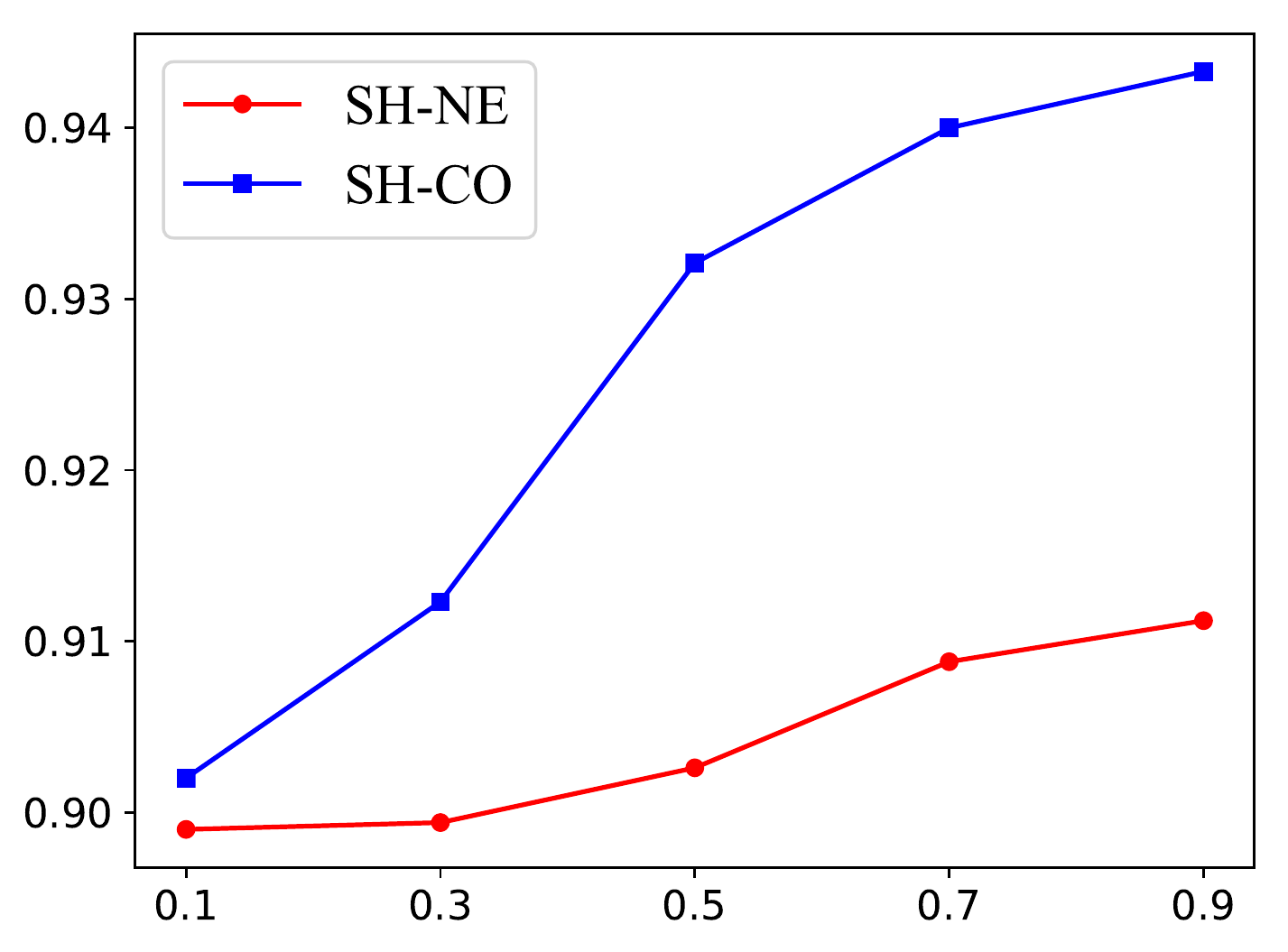}}
	\hspace{0.1in}
	\subfloat[MAP]{\includegraphics[width = 1.6in]{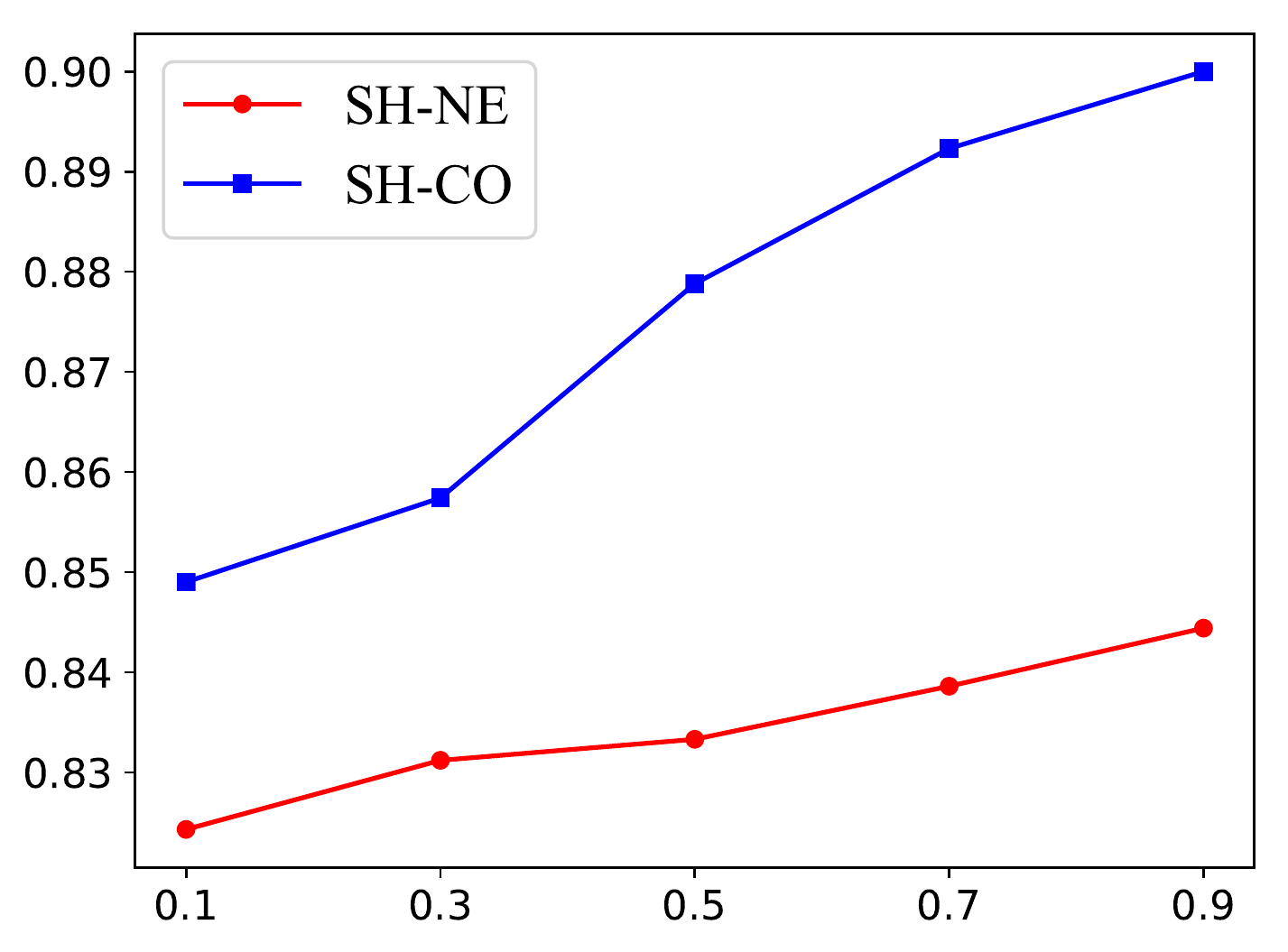}} 
	\caption{Our model performance on all metrics under different graph sparsity scales. X-axis is the ratio of links preserved in the sparse graph. Y-axis is related metric score.}
	\label{fig:sparsenss}
	\vspace{-1em} 
\end{figure}
To explore our model robustness for solving cold-start problem on sparse graphs, we randomly keep $\delta$ ratio of the total links in the sparse graph, where  $\delta=\{0.1,0.3,0.5,0.7,0.9\}$. Our model is trained on the whole main graph and each truncated sparse graph. The reported evaluation metrics under all sparse graphs with varied scales are visualized in Figure \ref{fig:sparsenss}. All metrics in both SH-NE and SH-CO datasets are in a rising trend with more links preserved in related sparse graphs. Compared with SH-NE dataset, SH-CO can benefit more from the sparse graph as it has a larger increase in all reported metrics. 

With more links are preserved, all evaluation metrics first grows rapidly. While the increasing speed slows down when 70\%-90\% links are preserved. This phenomenon can be seen in almost all plots in Figure \ref{fig:sparsenss}. It can be explained by the law of diminishing marginal utility, meaning that the marginal utility to bring new information derived from each additional link is declined.

Although classification related  metrics significantly increase with more links preserved (i.e., F1 score in SH-CO dataset increases 10\% when involving sparse graph), the retrieval related metrics (MRR, NDCG and MAP) do not change much in the mean time. One possible reason is that the information from the main shopping graph already contains enough information for the pairwise ranking in user triplets. For example, without considering sparse graph information, our model has already achieved high MRR score as 0.87 in SH-NE dataset and 0.89 in SH-CO dataset, which is already better than most baseline results running on the combined graph.

\subsection{Case Study}
\begin{figure}  
 \centering
 \includegraphics[width=1\columnwidth]{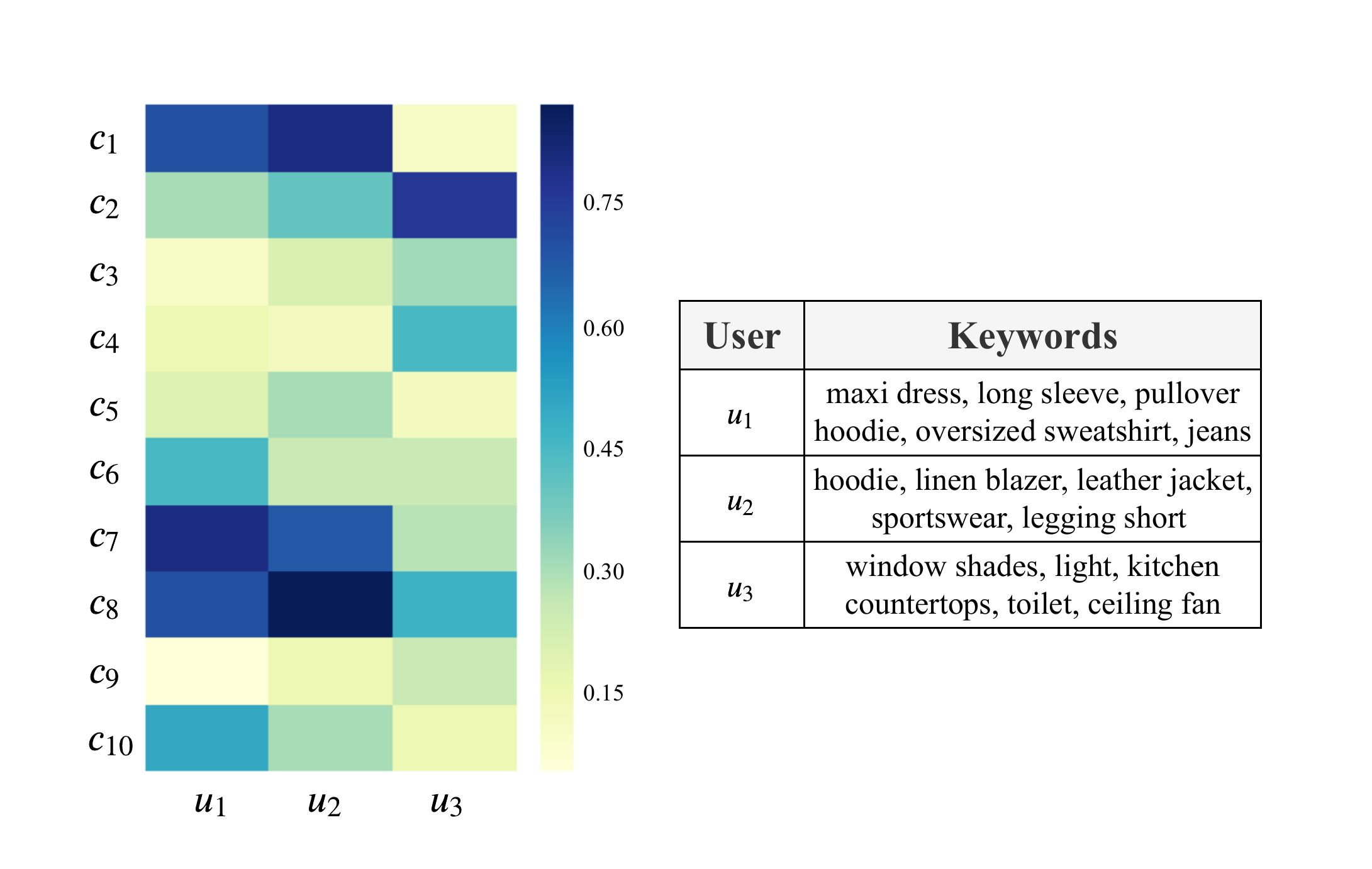}
  \vspace{-3em}
 \caption{The left part shows ten community affiliation scores of three selected users in the shopping graph. The right part shows the keywords of their purchased products.}
 
 \label{fig:case}
\end{figure}

In order to testify whether our proposed Community Recurrent Unit is able to accurately calculate user affiliation scores in each community, we choose three users (labelled from user $u_1$ to user $u_3$) and calculate their affiliation scores of ten selected communities (labelled from community $c_1$ to community $c_{10}$) in the main shopping graph. The detailed result is visualized in Figure \ref{fig:case}. In the left part of the figure, darker color indicates higher affiliation scores in the range from 0 to 1. Besides, in the shopping graph, we also extract all products purchased by each user, and manually select the most representative ones summarized as keywords in a table, which is demonstrated in the right part of Figure \ref{fig:case}.

From the right-part table, $u_1$ and $u_2$ share similar shopping interests in clothing products. $u_2$ is more like a girl fond of sports. And $u_1$ more tends to be a fashion girl. While the shopping interests of $u_3$ is much far away from the other users. All purchased products by $u_3$ is furnishing stuffs. It seems s/he is decorating her/his living space. Referring to the left part of Figure \ref{fig:case}, the affiliation distribution among ten communities between $u_1$ and $u_2$ are very similar. They both have a high weight in community $c_1$, $c_7$ and $c_8$. While the community distribution of $u_3$ is totally different. $u_3$ is heavily affiliated with community $c_2$. The results of our calculated community affiliation distribution is consistent with actual user behaviors, which indicates our Community Recurrent Unit (CRU) is  functional well to reflect user community information.

%% file: content/review.tex
\section{Related Works}

%
%
%
%
%

Exploring community structure in graphs has long been a fundamental research task in network science \cite{wang2017community}. It aims to group densely connected nodes in the same community while setting weakly connected nodes apart. Modularity-based approaches explore the community structure with the largest predefined modularity score. Spectral approaches either extract the main components from graph Laplacians or cut the whole graph into several disjoint chunks.   Dynamic approaches mostly detect  communities from random walks by maximizing its probability or minimizing the walk path description length \cite{ballal2020inference}. Recently, a surge of deep learning techniques are established to either learn node embeddings \cite{jiang2020task,gao2019edge2vec} for spectral clustering, or propose novel modules to detect communities in an end-to-end fashion \cite{luo2020deep}. Some other approaches also focus on special scenarios, i.e. personalized community detection \cite{gao2019efficient,gao2017personalized} or community-based information diffusion \cite{liu2016comparing}.

However, conventional approaches are solely dependent on the node connections in a single graph \cite{fortunato2016community}, which is vulnerable when the graph connectivity is sparse. To cope with this problem, \cite{amini2013pseudo} perturbs  sparse graphs by adding additional weak edges to connect
isolated components. \cite{guedon2016community,montanari2016semidefinite} both argue to calculate Grothendieck’s inequality on sparse graphs. \cite{krzakala2013spectral} adds a nonbacktracking walk on the directed
edges of sparse graphs to enhance existing spectral method performances. Similarly, \cite{chin2015stochastic} proposes a refined spectral method by adding an optimal recovery rate. Beyond adding subtle components into existing methods, two other types of approach can directly address sparse graph community detection problem by cross-graph learning (involving auxiliary information from other graphs) and pairwise learning (predicting the pairwise community relationship between nodes). 
 
By jointly training on multi-graphs, mutual pattern can be shared across graphs  \cite{hsu2017learning}. \cite{cheng2013flexible} uses a non-negative matrix factorization with a designed co-regularized penalty on multiple graphs to manipulate their communities simultaneously. \cite{long2014transfer} introduces a unified model to jointly maximize the empirical likelihood and preserves the geometric structure in  multi-graphs. \cite{chang2017cross} calibrates domain-specific graph Laplacians into a unified kernel, which detects graph patterns in semi-supervised fashion. \cite{wang2018cross} introduces a matrix factorization approach on two bipartite graphs simultaneously to measure the similarity between their shared nodes. \cite{liu2019jscn} operates
multi-layer spectral convolutions on different graphs to learn node communities. \cite{farseev2017cross} proposes a  regularized spectral clustering method to perform an efficient partitioning on multi-graphs. \cite{papalexakis2013more} introduces two different solutions to detect communities on multi-graphs based on minimum description length and tensor based decomposition principles.

Pairwise community detection offers an alternative view to avoid graph sparsity. Unlike other semi-supervised approaches \cite{fogel2018clustering,yu2018clustering} which just regard node pairwise relationships as extra constraints for whole graph community detection, the pairwise community relationship between nodes is the predicting target. However, as a subtle track of community detection, it is neglected and starts to attract research attention until recently. Among one of the few works, \cite{shukla2018clusternet} calculates the pairwise regularized KL-divergence from few labeled nodes, which is utilized to exploit the pairwise community relationship between abundant unlabeled data via convolutional autoencoder. \cite{ibrahim2019detecting,fogel2018clustering} extracts pairwise constraints using a mutual K-nearest neighbors approach to detect node pairwise community relationships.  
\cite{li2018simultaneous} leverages a multi-task to detect graph community and rank node pairwise community relationship simultaneously.  \cite{chaudhary2020community} considers the pairwise proximity among the nodes of the complex networks and unveils community structure based on that. \cite{chattopadhyay2020towards} uses a well-defined pairwise node similarity measure to identifies user communities in e-commerce.




%
%

%% file: content/conclusion.tex
\vspace{-0.5em}
\section{Conclusion}
Internet ecosystem presents star-shaped topology in the recent years: giant service providers, like Facebook, Google and Amazon, provide easy login channels to thousands of sites/apps, and users don't need experience laborious account creation anymore. In this study, motivated by this phenomena, we proposed a novel cross-graph community detection problem, which aims to detect user communities in sparse graphs by leveraging cross-graph information and pairwise learning techniques. It can substantially benefit small businesses or services to identify user groups and understand their interests. Extensive experiments on two real datasets demonstrate the superiority of our model comparing with all baselines. The limitation of this study is that we need to empirically detect communities to generate pseudo labels for training in a separate step. In the future, this process needs to be either omitted or merged to learn in an end-to-end manner. We also plan to upgrade our model by involving more auxiliary information, e.g., user/product textual profiles. Besides, Current model only considers information propagation between two graphs. A followup work needs to allow it run simultaneously among multiple graphs with appropriate graph-level filters.  

%% file: sample-sigconf.bbl

\begin{thebibliography}{40}


\ifx \showCODEN    \undefined \def \showCODEN     #1{\unskip}     \fi
\ifx \showDOI      \undefined \def \showDOI       #1{#1}\fi
\ifx \showISBNx    \undefined \def \showISBNx     #1{\unskip}     \fi
\ifx \showISBNxiii \undefined \def \showISBNxiii  #1{\unskip}     \fi
\ifx \showISSN     \undefined \def \showISSN      #1{\unskip}     \fi
\ifx \showLCCN     \undefined \def \showLCCN      #1{\unskip}     \fi
\ifx \shownote     \undefined \def \shownote      #1{#1}          \fi
\ifx \showarticletitle \undefined \def \showarticletitle #1{#1}   \fi
\ifx \showURL      \undefined \def \showURL       {\relax}        \fi
\providecommand\bibfield[2]{#2}
\providecommand\bibinfo[2]{#2}
\providecommand\natexlab[1]{#1}
\providecommand\showeprint[2][]{arXiv:#2}

\bibitem[\protect\citeauthoryear{Amini, Chen, Bickel, Levina,
  et~al\mbox{.}}{Amini et~al\mbox{.}}{2013}]%
        {amini2013pseudo}
\bibfield{author}{\bibinfo{person}{Arash~A Amini}, \bibinfo{person}{Aiyou
  Chen}, \bibinfo{person}{Peter~J Bickel}, \bibinfo{person}{Elizaveta Levina},
  {et~al\mbox{.}}} \bibinfo{year}{2013}\natexlab{}.
\newblock \showarticletitle{Pseudo-likelihood methods for community detection
  in large sparse networks}.
\newblock \bibinfo{journal}{\emph{The Annals of Statistics}}
  \bibinfo{volume}{41}, \bibinfo{number}{4} (\bibinfo{year}{2013}),
  \bibinfo{pages}{2097--2122}.
\newblock


\bibitem[\protect\citeauthoryear{Ballal, Kion-Crosby, and Morozov}{Ballal
  et~al\mbox{.}}{2020}]%
        {ballal2020inference}
\bibfield{author}{\bibinfo{person}{Aditya Ballal}, \bibinfo{person}{Willow
  Kion-Crosby}, {and} \bibinfo{person}{Alexandre Morozov}.}
  \bibinfo{year}{2020}\natexlab{}.
\newblock \showarticletitle{Inference of Network Communities using Random
  Walks}.
\newblock \bibinfo{journal}{\emph{Bulletin of the American Physical Society}}
  (\bibinfo{year}{2020}).
\newblock


\bibitem[\protect\citeauthoryear{Burges}{Burges}{2010}]%
        {burges2010ranknet}
\bibfield{author}{\bibinfo{person}{Christopher~JC Burges}.}
  \bibinfo{year}{2010}\natexlab{}.
\newblock \showarticletitle{From ranknet to lambdarank to lambdamart: An
  overview}.
\newblock \bibinfo{journal}{\emph{Learning}} \bibinfo{volume}{11},
  \bibinfo{number}{23-581} (\bibinfo{year}{2010}), \bibinfo{pages}{81}.
\newblock


\bibitem[\protect\citeauthoryear{Castrillo, Le{\'o}n, and G{\'o}mez}{Castrillo
  et~al\mbox{.}}{2017}]%
        {castrillo2017fast}
\bibfield{author}{\bibinfo{person}{Eduar Castrillo}, \bibinfo{person}{Elizabeth
  Le{\'o}n}, {and} \bibinfo{person}{Jonatan G{\'o}mez}.}
  \bibinfo{year}{2017}\natexlab{}.
\newblock \showarticletitle{Fast Heuristic Algorithm for Multi-scale
  Hierarchical Community Detection}. In \bibinfo{booktitle}{\emph{Proceedings
  of the 2017 IEEE/ACM International Conference on Advances in Social Networks
  Analysis and Mining 2017}}. ACM, \bibinfo{pages}{982--989}.
\newblock


\bibitem[\protect\citeauthoryear{Chamberlain, Levy-Kramer, Humby, and
  Deisenroth}{Chamberlain et~al\mbox{.}}{2018}]%
        {chamberlain2018real}
\bibfield{author}{\bibinfo{person}{Benjamin~Paul Chamberlain},
  \bibinfo{person}{Josh Levy-Kramer}, \bibinfo{person}{Clive Humby}, {and}
  \bibinfo{person}{Marc~Peter Deisenroth}.} \bibinfo{year}{2018}\natexlab{}.
\newblock \showarticletitle{Real-time community detection in full social
  networks on a laptop}.
\newblock \bibinfo{journal}{\emph{PloS one}} \bibinfo{volume}{13},
  \bibinfo{number}{1} (\bibinfo{year}{2018}), \bibinfo{pages}{e0188702}.
\newblock


\bibitem[\protect\citeauthoryear{Chang, Wu, Liu, and Yang}{Chang
  et~al\mbox{.}}{2017}]%
        {chang2017cross}
\bibfield{author}{\bibinfo{person}{Wei-Cheng Chang}, \bibinfo{person}{Yuexin
  Wu}, \bibinfo{person}{Hanxiao Liu}, {and} \bibinfo{person}{Yiming Yang}.}
  \bibinfo{year}{2017}\natexlab{}.
\newblock \showarticletitle{Cross-Domain Kernel Induction for Transfer
  Learning.}. In \bibinfo{booktitle}{\emph{AAAI}}. \bibinfo{pages}{1763--1769}.
\newblock


\bibitem[\protect\citeauthoryear{Chattopadhyay, Basu, Das, Ghosh, and
  Murthy}{Chattopadhyay et~al\mbox{.}}{2020}]%
        {chattopadhyay2020towards}
\bibfield{author}{\bibinfo{person}{Swarup Chattopadhyay},
  \bibinfo{person}{Tanmay Basu}, \bibinfo{person}{Asit~K Das},
  \bibinfo{person}{Kuntal Ghosh}, {and} \bibinfo{person}{Late~CA Murthy}.}
  \bibinfo{year}{2020}\natexlab{}.
\newblock \showarticletitle{Towards effective discovery of natural communities
  in complex networks and implications in e-commerce}.
\newblock \bibinfo{journal}{\emph{Electronic Commerce Research}}
  (\bibinfo{year}{2020}), \bibinfo{pages}{1--38}.
\newblock


\bibitem[\protect\citeauthoryear{Chaudhary and Singh}{Chaudhary and
  Singh}{2020}]%
        {chaudhary2020community}
\bibfield{author}{\bibinfo{person}{Laxmi Chaudhary} {and}
  \bibinfo{person}{Buddha Singh}.} \bibinfo{year}{2020}\natexlab{}.
\newblock \showarticletitle{Community detection using maximizing modularity and
  similarity measures in social networks}.
\newblock In \bibinfo{booktitle}{\emph{Smart Systems and IoT: Innovations in
  Computing}}. \bibinfo{publisher}{Springer}, \bibinfo{pages}{197--206}.
\newblock


\bibitem[\protect\citeauthoryear{Cheng, Zhang, Guo, Wu, Sullivan, and
  Wang}{Cheng et~al\mbox{.}}{2013}]%
        {cheng2013flexible}
\bibfield{author}{\bibinfo{person}{Wei Cheng}, \bibinfo{person}{Xiang Zhang},
  \bibinfo{person}{Zhishan Guo}, \bibinfo{person}{Yubao Wu},
  \bibinfo{person}{Patrick~F Sullivan}, {and} \bibinfo{person}{Wei Wang}.}
  \bibinfo{year}{2013}\natexlab{}.
\newblock \showarticletitle{Flexible and robust co-regularized multi-domain
  graph clustering}. In \bibinfo{booktitle}{\emph{Proceedings of the 19th ACM
  SIGKDD international conference on Knowledge discovery and data mining}}.
  ACM, \bibinfo{pages}{320--328}.
\newblock


\bibitem[\protect\citeauthoryear{Chin, Rao, and Vu}{Chin et~al\mbox{.}}{2015}]%
        {chin2015stochastic}
\bibfield{author}{\bibinfo{person}{Peter Chin}, \bibinfo{person}{Anup Rao},
  {and} \bibinfo{person}{Van Vu}.} \bibinfo{year}{2015}\natexlab{}.
\newblock \showarticletitle{Stochastic block model and community detection in
  sparse graphs: A spectral algorithm with optimal rate of recovery}. In
  \bibinfo{booktitle}{\emph{Conference on Learning Theory}}.
  \bibinfo{pages}{391--423}.
\newblock


\bibitem[\protect\citeauthoryear{Emmons and Mucha}{Emmons and Mucha}{2019}]%
        {emmons2019map}
\bibfield{author}{\bibinfo{person}{Scott Emmons} {and} \bibinfo{person}{Peter~J
  Mucha}.} \bibinfo{year}{2019}\natexlab{}.
\newblock \showarticletitle{Map equation with metadata: Varying the role of
  attributes in community detection}.
\newblock \bibinfo{journal}{\emph{Physical Review E}} \bibinfo{volume}{100},
  \bibinfo{number}{2} (\bibinfo{year}{2019}), \bibinfo{pages}{022301}.
\newblock


\bibitem[\protect\citeauthoryear{Farseev, Samborskii, Filchenkov, and
  Chua}{Farseev et~al\mbox{.}}{2017}]%
        {farseev2017cross}
\bibfield{author}{\bibinfo{person}{Aleksandr Farseev}, \bibinfo{person}{Ivan
  Samborskii}, \bibinfo{person}{Andrey Filchenkov}, {and}
  \bibinfo{person}{Tat-Seng Chua}.} \bibinfo{year}{2017}\natexlab{}.
\newblock \showarticletitle{Cross-domain recommendation via clustering on
  multi-layer graphs}. In \bibinfo{booktitle}{\emph{Proceedings of the 40th
  International ACM SIGIR Conference on Research and Development in Information
  Retrieval}}. ACM, \bibinfo{pages}{195--204}.
\newblock


\bibitem[\protect\citeauthoryear{Fogel, Averbuch-Elor, Goldberger, and
  Cohen-Or}{Fogel et~al\mbox{.}}{2018}]%
        {fogel2018clustering}
\bibfield{author}{\bibinfo{person}{Sharon Fogel}, \bibinfo{person}{Hadar
  Averbuch-Elor}, \bibinfo{person}{Jacov Goldberger}, {and}
  \bibinfo{person}{Daniel Cohen-Or}.} \bibinfo{year}{2018}\natexlab{}.
\newblock \showarticletitle{Clustering-driven Deep Embedding with Pairwise
  Constraints}.
\newblock \bibinfo{journal}{\emph{arXiv preprint arXiv:1803.08457}}
  (\bibinfo{year}{2018}).
\newblock


\bibitem[\protect\citeauthoryear{Fortunato and Hric}{Fortunato and
  Hric}{2016}]%
        {fortunato2016community}
\bibfield{author}{\bibinfo{person}{Santo Fortunato} {and}
  \bibinfo{person}{Darko Hric}.} \bibinfo{year}{2016}\natexlab{}.
\newblock \showarticletitle{Community detection in networks: A user guide}.
\newblock \bibinfo{journal}{\emph{Physics Reports}}  \bibinfo{volume}{659}
  (\bibinfo{year}{2016}), \bibinfo{pages}{1--44}.
\newblock


\bibitem[\protect\citeauthoryear{Gao, Fu, Ouyang, Tsutsui, Liu, Yang, Gessner,
  Foote, Wild, Ding, et~al\mbox{.}}{Gao et~al\mbox{.}}{2019a}]%
        {gao2019edge2vec}
\bibfield{author}{\bibinfo{person}{Zheng Gao}, \bibinfo{person}{Gang Fu},
  \bibinfo{person}{Chunping Ouyang}, \bibinfo{person}{Satoshi Tsutsui},
  \bibinfo{person}{Xiaozhong Liu}, \bibinfo{person}{Jeremy Yang},
  \bibinfo{person}{Christopher Gessner}, \bibinfo{person}{Brian Foote},
  \bibinfo{person}{David Wild}, \bibinfo{person}{Ying Ding}, {et~al\mbox{.}}}
  \bibinfo{year}{2019}\natexlab{a}.
\newblock \showarticletitle{edge2vec: Representation learning using edge
  semantics for biomedical knowledge discovery}.
\newblock \bibinfo{journal}{\emph{BMC bioinformatics}} \bibinfo{volume}{20},
  \bibinfo{number}{1} (\bibinfo{year}{2019}), \bibinfo{pages}{306}.
\newblock


\bibitem[\protect\citeauthoryear{Gao, Guo, and Liu}{Gao et~al\mbox{.}}{2019b}]%
        {gao2019efficient}
\bibfield{author}{\bibinfo{person}{Zheng Gao}, \bibinfo{person}{Chun Guo},
  {and} \bibinfo{person}{Xiaozhong Liu}.} \bibinfo{year}{2019}\natexlab{b}.
\newblock \showarticletitle{Efficient personalized community detection via
  genetic evolution}. In \bibinfo{booktitle}{\emph{Proceedings of the Genetic
  and Evolutionary Computation Conference}}. \bibinfo{pages}{383--391}.
\newblock


\bibitem[\protect\citeauthoryear{Gao and Liu}{Gao and Liu}{2017}]%
        {gao2017personalized}
\bibfield{author}{\bibinfo{person}{Zheng Gao} {and} \bibinfo{person}{Xiaozhong
  Liu}.} \bibinfo{year}{2017}\natexlab{}.
\newblock \showarticletitle{Personalized community detection in scholarly
  network}.
\newblock \bibinfo{journal}{\emph{iConference 2017 Proceedings Vol. 2}}
  (\bibinfo{year}{2017}).
\newblock


\bibitem[\protect\citeauthoryear{Grover and Leskovec}{Grover and
  Leskovec}{2016}]%
        {grover2016node2vec}
\bibfield{author}{\bibinfo{person}{Aditya Grover} {and} \bibinfo{person}{Jure
  Leskovec}.} \bibinfo{year}{2016}\natexlab{}.
\newblock \showarticletitle{node2vec: Scalable feature learning for networks}.
  In \bibinfo{booktitle}{\emph{Proceedings of the 22nd ACM SIGKDD international
  conference on Knowledge discovery and data mining}}. ACM,
  \bibinfo{pages}{855--864}.
\newblock


\bibitem[\protect\citeauthoryear{Gu{\'e}don and Vershynin}{Gu{\'e}don and
  Vershynin}{2016}]%
        {guedon2016community}
\bibfield{author}{\bibinfo{person}{Olivier Gu{\'e}don} {and}
  \bibinfo{person}{Roman Vershynin}.} \bibinfo{year}{2016}\natexlab{}.
\newblock \showarticletitle{Community detection in sparse networks via
  Grothendieck’s inequality}.
\newblock \bibinfo{journal}{\emph{Probability Theory and Related Fields}}
  \bibinfo{volume}{165}, \bibinfo{number}{3-4} (\bibinfo{year}{2016}),
  \bibinfo{pages}{1025--1049}.
\newblock


\bibitem[\protect\citeauthoryear{Haq, Moradi, and Wang}{Haq
  et~al\mbox{.}}{2019}]%
        {haq2019community}
\bibfield{author}{\bibinfo{person}{Nandinee~Fariah Haq}, \bibinfo{person}{Mehdi
  Moradi}, {and} \bibinfo{person}{Z~Jane Wang}.}
  \bibinfo{year}{2019}\natexlab{}.
\newblock \showarticletitle{Community structure detection from networks with
  weighted modularity}.
\newblock \bibinfo{journal}{\emph{Pattern Recognition Letters}}
  \bibinfo{volume}{122} (\bibinfo{year}{2019}), \bibinfo{pages}{14--22}.
\newblock


\bibitem[\protect\citeauthoryear{Hsu, Lv, and Kira}{Hsu et~al\mbox{.}}{2017}]%
        {hsu2017learning}
\bibfield{author}{\bibinfo{person}{Yen-Chang Hsu}, \bibinfo{person}{Zhaoyang
  Lv}, {and} \bibinfo{person}{Zsolt Kira}.} \bibinfo{year}{2017}\natexlab{}.
\newblock \showarticletitle{Learning to cluster in order to Transfer across
  domains and tasks}.
\newblock \bibinfo{journal}{\emph{arXiv preprint arXiv:1711.10125}}
  (\bibinfo{year}{2017}).
\newblock


\bibitem[\protect\citeauthoryear{Ibrahim, Missaoui, and Messaoudi}{Ibrahim
  et~al\mbox{.}}{2019}]%
        {ibrahim2019detecting}
\bibfield{author}{\bibinfo{person}{Mohamed-Hamza Ibrahim},
  \bibinfo{person}{Rokia Missaoui}, {and} \bibinfo{person}{Abir Messaoudi}.}
  \bibinfo{year}{2019}\natexlab{}.
\newblock \showarticletitle{Detecting Local Community Structures in Social
  Networks Using Concept Interestingness}.
\newblock \bibinfo{journal}{\emph{arXiv preprint arXiv:1902.03109}}
  (\bibinfo{year}{2019}).
\newblock


\bibitem[\protect\citeauthoryear{Jiang, Gao, Lan, Yang, Lu, and Liu}{Jiang
  et~al\mbox{.}}{2020}]%
        {jiang2020task}
\bibfield{author}{\bibinfo{person}{Zhuoren Jiang}, \bibinfo{person}{Zheng Gao},
  \bibinfo{person}{Jinjiong Lan}, \bibinfo{person}{Hongxia Yang},
  \bibinfo{person}{Yao Lu}, {and} \bibinfo{person}{Xiaozhong Liu}.}
  \bibinfo{year}{2020}\natexlab{}.
\newblock \showarticletitle{Task-Oriented Genetic Activation for Large-Scale
  Complex Heterogeneous Graph Embedding}. In
  \bibinfo{booktitle}{\emph{Proceedings of The Web Conference 2020}}.
  \bibinfo{pages}{1581--1591}.
\newblock


\bibitem[\protect\citeauthoryear{Kheirkhahzadeh, Lancichinetti, and
  Rosvall}{Kheirkhahzadeh et~al\mbox{.}}{2016}]%
        {kheirkhahzadeh2016efficient}
\bibfield{author}{\bibinfo{person}{Masoumeh Kheirkhahzadeh},
  \bibinfo{person}{Andrea Lancichinetti}, {and} \bibinfo{person}{Martin
  Rosvall}.} \bibinfo{year}{2016}\natexlab{}.
\newblock \showarticletitle{Efficient community detection of network flows for
  varying Markov times and bipartite networks}.
\newblock \bibinfo{journal}{\emph{Physical Review E}} \bibinfo{volume}{93},
  \bibinfo{number}{3} (\bibinfo{year}{2016}), \bibinfo{pages}{032309}.
\newblock


\bibitem[\protect\citeauthoryear{Krzakala, Moore, Mossel, Neeman, Sly,
  Zdeborov{\'a}, and Zhang}{Krzakala et~al\mbox{.}}{2013}]%
        {krzakala2013spectral}
\bibfield{author}{\bibinfo{person}{Florent Krzakala},
  \bibinfo{person}{Cristopher Moore}, \bibinfo{person}{Elchanan Mossel},
  \bibinfo{person}{Joe Neeman}, \bibinfo{person}{Allan Sly},
  \bibinfo{person}{Lenka Zdeborov{\'a}}, {and} \bibinfo{person}{Pan Zhang}.}
  \bibinfo{year}{2013}\natexlab{}.
\newblock \showarticletitle{Spectral redemption in clustering sparse networks}.
\newblock \bibinfo{journal}{\emph{Proceedings of the National Academy of
  Sciences}} \bibinfo{volume}{110}, \bibinfo{number}{52}
  (\bibinfo{year}{2013}), \bibinfo{pages}{20935--20940}.
\newblock


\bibitem[\protect\citeauthoryear{Li, Baba, and Kashima}{Li
  et~al\mbox{.}}{2018}]%
        {li2018simultaneous}
\bibfield{author}{\bibinfo{person}{Jiyi Li}, \bibinfo{person}{Yukino Baba},
  {and} \bibinfo{person}{Hisashi Kashima}.} \bibinfo{year}{2018}\natexlab{}.
\newblock \showarticletitle{Simultaneous Clustering and Ranking from Pairwise
  Comparisons.}. In \bibinfo{booktitle}{\emph{IJCAI}}.
  \bibinfo{pages}{1554--1560}.
\newblock


\bibitem[\protect\citeauthoryear{Liu, Yu, Gao, Xia, and Bollen}{Liu
  et~al\mbox{.}}{2016}]%
        {liu2016comparing}
\bibfield{author}{\bibinfo{person}{Xiaozhong Liu}, \bibinfo{person}{Xing Yu},
  \bibinfo{person}{Zheng Gao}, \bibinfo{person}{Tian Xia}, {and}
  \bibinfo{person}{Johan Bollen}.} \bibinfo{year}{2016}\natexlab{}.
\newblock \showarticletitle{Comparing community-based information adoption and
  diffusion across different microblogging sites}. In
  \bibinfo{booktitle}{\emph{Proceedings of the 27th ACM Conference on Hypertext
  and Social Media}}. \bibinfo{pages}{103--112}.
\newblock


\bibitem[\protect\citeauthoryear{Liu, Zheng, Zhang, Han, and Yu}{Liu
  et~al\mbox{.}}{2019}]%
        {liu2019jscn}
\bibfield{author}{\bibinfo{person}{Zhiwei Liu}, \bibinfo{person}{Lei Zheng},
  \bibinfo{person}{Jiawei Zhang}, \bibinfo{person}{Jiayu Han}, {and}
  \bibinfo{person}{Philip~S Yu}.} \bibinfo{year}{2019}\natexlab{}.
\newblock \showarticletitle{JSCN: Joint Spectral Convolutional Network for
  Cross Domain Recommendation}.
\newblock \bibinfo{journal}{\emph{arXiv preprint arXiv:1910.08219}}
  (\bibinfo{year}{2019}).
\newblock


\bibitem[\protect\citeauthoryear{Long, Wang, Ding, Shen, and Yang}{Long
  et~al\mbox{.}}{2014}]%
        {long2014transfer}
\bibfield{author}{\bibinfo{person}{Mingsheng Long}, \bibinfo{person}{Jianmin
  Wang}, \bibinfo{person}{Guiguang Ding}, \bibinfo{person}{Dou Shen}, {and}
  \bibinfo{person}{Qiang Yang}.} \bibinfo{year}{2014}\natexlab{}.
\newblock \showarticletitle{Transfer Learning with Graph Co-Regularization.}
\newblock \bibinfo{journal}{\emph{IEEE Trans. Knowl. Data Eng.}}
  \bibinfo{volume}{26}, \bibinfo{number}{7} (\bibinfo{year}{2014}),
  \bibinfo{pages}{1805--1818}.
\newblock


\bibitem[\protect\citeauthoryear{Luo, Ni, Wang, Bian, Yu, and Zhang}{Luo
  et~al\mbox{.}}{2020}]%
        {luo2020deep}
\bibfield{author}{\bibinfo{person}{Dongsheng Luo}, \bibinfo{person}{Jingchao
  Ni}, \bibinfo{person}{Suhang Wang}, \bibinfo{person}{Yuchen Bian},
  \bibinfo{person}{Xiong Yu}, {and} \bibinfo{person}{Xiang Zhang}.}
  \bibinfo{year}{2020}\natexlab{}.
\newblock \showarticletitle{Deep Multi-Graph Clustering via Attentive
  Cross-Graph Association}. In \bibinfo{booktitle}{\emph{Proceedings of the
  13th International Conference on Web Search and Data Mining}}.
  \bibinfo{pages}{393--401}.
\newblock


\bibitem[\protect\citeauthoryear{Montanari and Sen}{Montanari and Sen}{2016}]%
        {montanari2016semidefinite}
\bibfield{author}{\bibinfo{person}{Andrea Montanari} {and}
  \bibinfo{person}{Subhabrata Sen}.} \bibinfo{year}{2016}\natexlab{}.
\newblock \showarticletitle{Semidefinite programs on sparse random graphs and
  their application to community detection}. In
  \bibinfo{booktitle}{\emph{Proceedings of the forty-eighth annual ACM
  symposium on Theory of Computing}}. \bibinfo{pages}{814--827}.
\newblock


\bibitem[\protect\citeauthoryear{Papalexakis, Akoglu, and Ience}{Papalexakis
  et~al\mbox{.}}{2013}]%
        {papalexakis2013more}
\bibfield{author}{\bibinfo{person}{Evangelos~E Papalexakis},
  \bibinfo{person}{Leman Akoglu}, {and} \bibinfo{person}{Dino Ience}.}
  \bibinfo{year}{2013}\natexlab{}.
\newblock \showarticletitle{Do more views of a graph help? community detection
  and clustering in multi-graphs}. In \bibinfo{booktitle}{\emph{Proceedings of
  the 16th International Conference on Information Fusion}}. IEEE,
  \bibinfo{pages}{899--905}.
\newblock


\bibitem[\protect\citeauthoryear{Shukla, Cheema, and Anand}{Shukla
  et~al\mbox{.}}{2018}]%
        {shukla2018clusternet}
\bibfield{author}{\bibinfo{person}{Ankita Shukla},
  \bibinfo{person}{Gullal~Singh Cheema}, {and} \bibinfo{person}{Saket Anand}.}
  \bibinfo{year}{2018}\natexlab{}.
\newblock \showarticletitle{ClusterNet: Semi-Supervised Clustering using Neural
  Networks}.
\newblock \bibinfo{journal}{\emph{arXiv preprint arXiv:1806.01547}}
  (\bibinfo{year}{2018}).
\newblock


\bibitem[\protect\citeauthoryear{Sun, Qu, Hoffmann, Huang, and Tang}{Sun
  et~al\mbox{.}}{2019}]%
        {sun2019vgraph}
\bibfield{author}{\bibinfo{person}{Fan-Yun Sun}, \bibinfo{person}{Meng Qu},
  \bibinfo{person}{Jordan Hoffmann}, \bibinfo{person}{Chin-Wei Huang}, {and}
  \bibinfo{person}{Jian Tang}.} \bibinfo{year}{2019}\natexlab{}.
\newblock \showarticletitle{vGraph: A Generative Model for Joint Community
  Detection and Node Representation Learning}.
\newblock \bibinfo{journal}{\emph{arXiv preprint arXiv:1906.07159}}
  (\bibinfo{year}{2019}).
\newblock


\bibitem[\protect\citeauthoryear{Tang, Marin, Ayed, and Boykov}{Tang
  et~al\mbox{.}}{2019}]%
        {tang2019kernel}
\bibfield{author}{\bibinfo{person}{Meng Tang}, \bibinfo{person}{Dmitrii Marin},
  \bibinfo{person}{Ismail~Ben Ayed}, {and} \bibinfo{person}{Yuri Boykov}.}
  \bibinfo{year}{2019}\natexlab{}.
\newblock \showarticletitle{Kernel cuts: Kernel and spectral clustering meet
  regularization}.
\newblock \bibinfo{journal}{\emph{International Journal of Computer Vision}}
  \bibinfo{volume}{127}, \bibinfo{number}{5} (\bibinfo{year}{2019}),
  \bibinfo{pages}{477--511}.
\newblock


\bibitem[\protect\citeauthoryear{Wang, Cui, Wang, Pei, Zhu, and Yang}{Wang
  et~al\mbox{.}}{2017}]%
        {wang2017community}
\bibfield{author}{\bibinfo{person}{Xiao Wang}, \bibinfo{person}{Peng Cui},
  \bibinfo{person}{Jing Wang}, \bibinfo{person}{Jian Pei},
  \bibinfo{person}{Wenwu Zhu}, {and} \bibinfo{person}{Shiqiang Yang}.}
  \bibinfo{year}{2017}\natexlab{}.
\newblock \showarticletitle{Community preserving network embedding}. In
  \bibinfo{booktitle}{\emph{Thirty-First AAAI Conference on Artificial
  Intelligence}}.
\newblock


\bibitem[\protect\citeauthoryear{Wang, Peng, Wang, Philip, Fu, and Hong}{Wang
  et~al\mbox{.}}{2018}]%
        {wang2018cross}
\bibfield{author}{\bibinfo{person}{Xinghua Wang}, \bibinfo{person}{Zhaohui
  Peng}, \bibinfo{person}{Senzhang Wang}, \bibinfo{person}{S~Yu Philip},
  \bibinfo{person}{Wenjing Fu}, {and} \bibinfo{person}{Xiaoguang Hong}.}
  \bibinfo{year}{2018}\natexlab{}.
\newblock \showarticletitle{Cross-domain recommendation for cold-start users
  via neighborhood based feature mapping}. In
  \bibinfo{booktitle}{\emph{International Conference on Database Systems for
  Advanced Applications}}. Springer, \bibinfo{pages}{158--165}.
\newblock


\bibitem[\protect\citeauthoryear{Yang and Leskovec}{Yang and Leskovec}{2013}]%
        {yang2013overlapping}
\bibfield{author}{\bibinfo{person}{Jaewon Yang} {and} \bibinfo{person}{Jure
  Leskovec}.} \bibinfo{year}{2013}\natexlab{}.
\newblock \showarticletitle{Overlapping community detection at scale: a
  nonnegative matrix factorization approach}. In
  \bibinfo{booktitle}{\emph{Proceedings of the sixth ACM international
  conference on Web search and data mining}}. ACM, \bibinfo{pages}{587--596}.
\newblock


\bibitem[\protect\citeauthoryear{Yang, Cao, He, Wang, Wang, and Zhang}{Yang
  et~al\mbox{.}}{2016}]%
        {yang2016modularity}
\bibfield{author}{\bibinfo{person}{Liang Yang}, \bibinfo{person}{Xiaochun Cao},
  \bibinfo{person}{Dongxiao He}, \bibinfo{person}{Chuan Wang},
  \bibinfo{person}{Xiao Wang}, {and} \bibinfo{person}{Weixiong Zhang}.}
  \bibinfo{year}{2016}\natexlab{}.
\newblock \showarticletitle{Modularity Based Community Detection with Deep
  Learning.}. In \bibinfo{booktitle}{\emph{IJCAI}}.
  \bibinfo{pages}{2252--2258}.
\newblock


\bibitem[\protect\citeauthoryear{Yu, Elhabian, and Whitaker}{Yu
  et~al\mbox{.}}{2018}]%
        {yu2018clustering}
\bibfield{author}{\bibinfo{person}{Yen-Yun Yu}, \bibinfo{person}{Shireen~Y
  Elhabian}, {and} \bibinfo{person}{Ross~T Whitaker}.}
  \bibinfo{year}{2018}\natexlab{}.
\newblock \showarticletitle{Clustering With Pairwise Relationships: A
  Generative Approach}.
\newblock \bibinfo{journal}{\emph{arXiv preprint arXiv:1805.02285}}
  (\bibinfo{year}{2018}).
\newblock


\end{thebibliography}
